\documentclass[10pt,twocolumn,preprintnumbers,nofootinbib,prd,superscriptaddress,aps]{revtex4-1}

\usepackage{graphics}      
\usepackage[utf8]{inputenc} 
\usepackage{graphicx,amssymb,amsmath,amsthm,amsfonts,epstopdf,epsfig,epsf,times}
\usepackage[linktocpage]{hyperref}
\usepackage[usenames]{color}
\usepackage{textcomp}
\usepackage{bm}
\usepackage{latexsym}
\usepackage{rotating}
\usepackage{hyperref}
\usepackage{color}
\usepackage{enumerate}
\usepackage{tensor}
\usepackage{stmaryrd}
\usepackage[normalem]{ulem}
\usepackage{mathtools}
\usepackage{url}
\usepackage{multirow}
\usepackage{graphicx}
\usepackage{mathtools}
\usepackage{verbatim}
\usepackage{soul,xcolor}
\usepackage{xspace}
\usepackage{amsmath}
\usepackage{xfrac}
\setstcolor{red}
\usepackage{esint}
\usepackage{physics}
\usepackage{subcaption}

\renewcommand{\vec}[1]{\boldsymbol{#1}}
\def\be{\begin{equation}}
\def\ee{\end{equation}}

\def\oc{\omega_c}
\def\ok{\omega_K}

\newcommand{\beq}{\begin{eqnarray}}

\newcommand{\eeq}{\end{eqnarray}} 
\newcommand{\submu}{_{\mu}}

\newcommand{\submunu}{_{\mu \nu}}
\newcommand{\supmu}{^{\mu}}
\newcommand{\supnu}{^{\nu}}

\newcommand{\del}{\partial}

\newcommand{\mE}{\mathcal{E}}
\newcommand{\mL}{\mathcal{L}}
\newcommand{\lr}[1]{\left( #1\right)}

\newcommand{\overbar}[1]{\mkern 1.5mu\overline{\mkern-1.5mu#1\mkern-1.5mu}\mkern 1.5mu}

\newcommand{\js}[1]{{\textcolor{orange}	{\sf{[JS: #1]}}}}
\newcommand{\jn}[1]{{\textcolor{red}	{\sf{[JN: #1]}}}}

\begin{document}

\title{Radiation reaction in weakly magnetized black holes: \\
can the tail term be ignored in the strong field regime?}
\author{João S. Santos}
\affiliation{CENTRA, Departamento de F\'{\i}sica, Instituto Superior T\'ecnico -- IST, Universidade de Lisboa -- UL,
Avenida Rovisco Pais 1, 1049-001 Lisboa, Portugal}
\author{Vitor Cardoso}
\affiliation{CENTRA, Departamento de F\'{\i}sica, Instituto Superior T\'ecnico -- IST, Universidade de Lisboa -- UL,
Avenida Rovisco Pais 1, 1049-001 Lisboa, Portugal}
\affiliation{Niels Bohr International Academy, Niels Bohr Institute, Blegdamsvej 17, 2100 Copenhagen, Denmark}
\author{José Natário}
\affiliation{CAMGSD, Departamento de Matem\'{a}tica, Instituto Superior T\'ecnico -- IST, Universidade de Lisboa -- UL,
Avenida Rovisco Pais 1, 1049-001 Lisboa, Portugal}

\date{\today}

\begin{abstract}
We study radiation from charged particles in circular motion around a Schwarzschild black hole immersed in an asymptotically uniform magnetic field. In curved space, the radiation reaction force is described by the DeWitt-Brehme equation, which includes a complicated, non-local tail term. We show that, contrary to some claims in the literature, this term cannot, in general, be neglected. 
We account for self-force effects directly by calculating the electromagnetic energy flux at infinity and on the horizon. The radiative field is obtained using black hole perturbation theory. 
We solve the relevant equations analytically, in the low-frequency and slow-motion approximation, as well as numerically in the general case. Our results show that great care must be taken when neglecting the tail term, which is often fundamental to capture the dynamics of the particle: in fact, it only seems to be negligible when the magnetic force greatly dominates the gravitational force, so that the motion is well described by the Abraham--Lorentz--Dirac equation. 
We also report a curious ``horizon dominance effect" that occurs for a radiating particle in a circular orbit around a black hole (emitting either scalar, electromagnetic or gravitational waves): for fixed orbital radius, the fraction of energy that is absorbed by the black hole can be made arbitrarily large by decreasing the particle velocity. 
\end{abstract}

\maketitle
\section{Introduction} \label{sec:intro}
%

The self-force (or radiation reaction) problem concerns the interaction of a particle with its own radiative field~\cite{Dirac:1938nz,DEWITT1960220,HOBBS1968141,Smith:1980tv,Mino:1996nk,Quinn:1996am,Poisson:1999tv,Gralla:2009md,Poisson:2011nh,Detweiler:2011tt,Barack:2018yvs,Pound:2009sm}. Understanding the self-force problem in General Relativity, in particular in the context of a particle orbiting a compact object like a black hole (BH), is instrumental for next generation gravitational wave detectors~\cite{LISAConsortiumWaveformWorkingGroup:2023arg}, and is a very active topic of research~\cite{Breuer:1973kt,Detweiler:1978ge,Galtsov:1982hwm,Poisson:1993vp,Detweiler:2002mi,Pound:2012nt, Torres:2020fye,PanossoMacedo:2022fdi, German:2023bye, Miller:2023ers,Bourg:2024vre}. 

Particles in the vicinity of BHs can attain ultra-relativistic velocities~\cite{Bardeen:1972fi}, and we can expect matter to be ionized, so it makes sense to consider particles with electric charge. These particles generate electromagnetic and gravitational fields, both of which give rise to self-force effects. In this work, we focus on electromagnetic radiation reaction, which has been less studied than its gravitational counterpart in the context of BH physics.

Astrophysical BHs are often surrounded by intense magnetic fields~\cite{2011AstBu..66..320P,Eatough:2013nva,Baczko:2016opl,Daly:2019srb, EventHorizonTelescope:2021srq}, which are supported by accretion disks of ionized matter. As such, we also want to include the interaction of the charged particle with a background magnetic field. We consider the simplest case of a Schwarzschild BH surrounded by an asymptotically uniform magnetic field. This system has been amply studied in the absence of radiation~\cite{Frolov:2010mi,Frolov:2011ea,Zahrani:2013up,Kolos:2015iva,Qi:2023brf,Baker:2023gdc}, with a few notable exceptions~\cite{Aliev:1980hz, Sokolov:1978tc, Galtsov:1982hwm, 2018ApJ...861....2T}.

For simplicity, we opt to include the magnetic field as a perturbative effect. In other words, we assume that its backreaction on the geometry is negligible. Such a magnetic field can be computed using the results in Ref.~\cite{PhysRevD.10.1680}. Treating the magnetic field perturbatively means that we must restrict ourselves to a region of the Schwarzschild spacetime satisfying~\cite{Galtsov:1978ag}
\begin{equation}
	\frac{r }{G M / c^2} \ll   4.7\times 10^{19} \lr{\frac{1\, \text{Gauss}}{B_0}}\lr{\frac{M_\odot}{M}} \, .\label{eq:limit_validity}
\end{equation}
If we replace the typical values of magnetic field $B_0\sim 10^4 -- 10^8$ Gauss~\cite{2011AstBu..66..320P,Eatough:2013nva,Baczko:2016opl,Daly:2019srb}, we find there is ample room to study this system within the weak magnetic field approximation.

The equation of motion of a radiating charged particle in a curved background is the De Witt--Brehme equation~\cite{DEWITT1960220,HOBBS1968141,Poisson:2011nh}. This equation contains a particularly complicated, non-local, tail term. Some of the reference works in the literature concerning radiation reaction in weakly magnetized BHs claim that the tail term can be neglected~\cite{2018ApJ...861....2T}. This results in the appearance of a counterintuitive ``orbital widening'' effect~\cite{Tursunov:2018udx,Juraev:2024dju}. 

In a previous publication, working under a weak field approximation, we showed that the tail term must be included, unless the Newtonian gravitational force is negligible in comparison with the Lorentz force~\cite{Santos:2023uka}. We also showed that there can be no ``orbital widening'' in this limit, consistently with previous studies~\cite{Gron:2008tr}. A generalization of the result to the strong field case was however missing, and is provided here. We conclude that the tail term must be included except when the motion is dominated by magnetic effects. In all cases, there is no ``orbital widening," at least for particles in circular orbit.


This work is not the first concerning radiation reaction in weakly magnetized BHs. Still, we present for the first time formulas describing the energy flux for arbitrary multipolar modes, in the slow-motion and low-frequency approximation, for any value of orbital frequency of the particle. Incidentally, we give a fully detailed description of the method to find solutions of the Teukolsky equation in that approximation, which seems to be missing in the literature (see Appendix~\ref{app:matched_asymptotic_expansions}). More importantly, we give a full numerical solution to the Teukolsky equation. In the low-frequency and slow motion regime our analytical and numerical results show excellent agreement, and establish without doubt that the tail term cannot be neglected. Finally, we report a ``horizon dominance effect" (already encountered in~\cite{Galtsov:1982hwm,Brito:2012gw}), which we characterize in detail and show to also occur in flat space for appropriate absorbing boundary conditions (see Appendix~\ref{app:flat_space}).

This work is organized as follows: In Sec.~\ref{sec:background} we give an overview of the setup we are studying and of the DeWitt-Brehme equation, namely the tail term. Then, in Sec.~\ref{sec:perturbations}, we show how the radiation field of a charged particle in circular orbit and the corresponding energy fluxes at infinity and on the BH horizon can be calculated using the Teukolsky equation. Using these results, we obtain in Sec.~\ref{sec:analytical} analytical formulas for the slow-motion and low-frequency approximation. Finally, in Sec.~\ref{sec:numerical} we obtain numerical results for the general case.

We use a system of units with $G=c=1$, and use Gaussian units for electromagnetism, meaning that we have $4\pi\varepsilon_0 = 1$ and $\mu_0 = 4 \pi$. The metric signature is $\left( -,+,+,+ \right)$, and Greek indices run from 0 to 3. 
%
%
%
\section{Setup} \label{sec:background}
%
%
\subsection{Weakly magnetized Schwarzschild black hole} \label{sec:mag_bh}
%
\begin{figure*}[ht!]
    \centering
    \hspace{0.07 \textwidth}
    \begin{subfigure}{0.4\textwidth}
         \centering
   	\includegraphics[width=\textwidth]{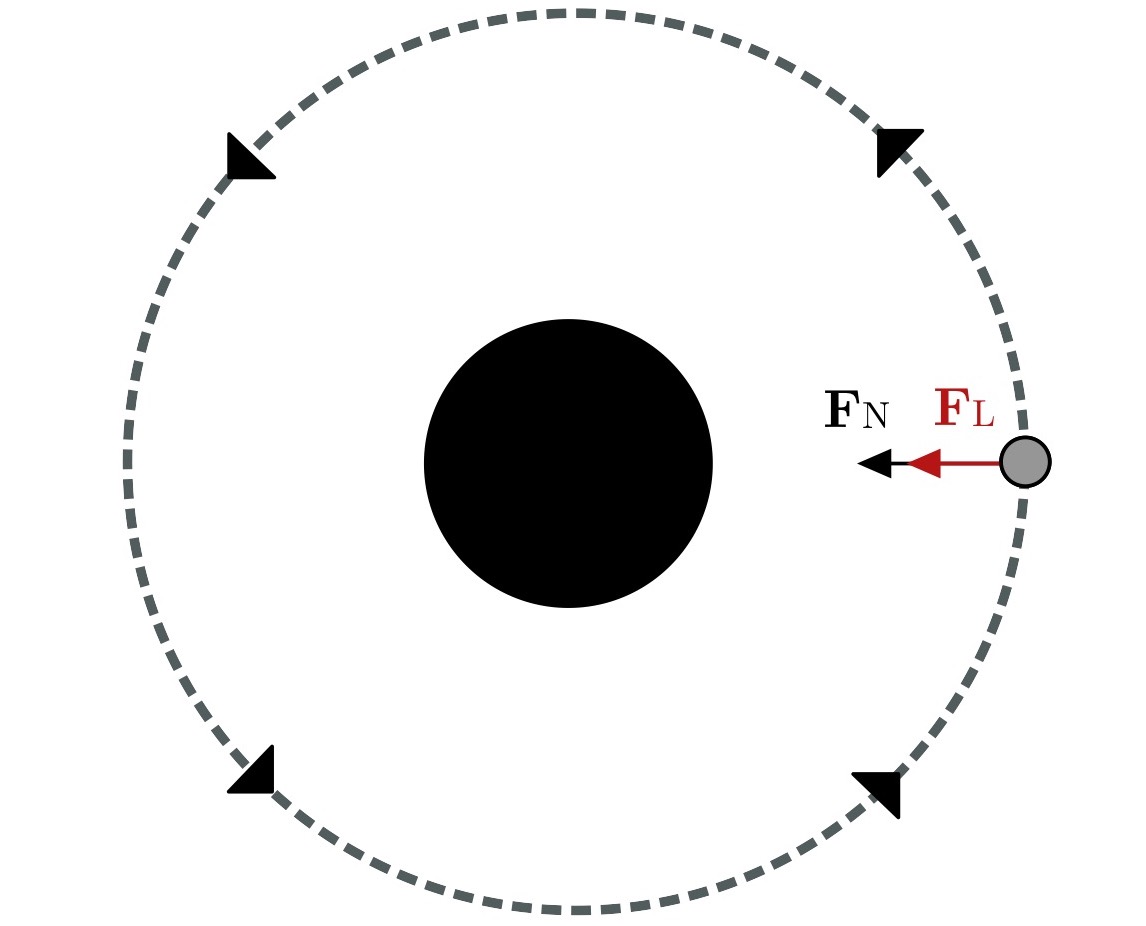}
         \caption{Minus configuration (MC): $\oc < 0$.}
         \label{fig:minus}
     \end{subfigure}
     \hfill
     \begin{subfigure}{0.4\textwidth}
         \centering
   	\includegraphics[width=\textwidth]{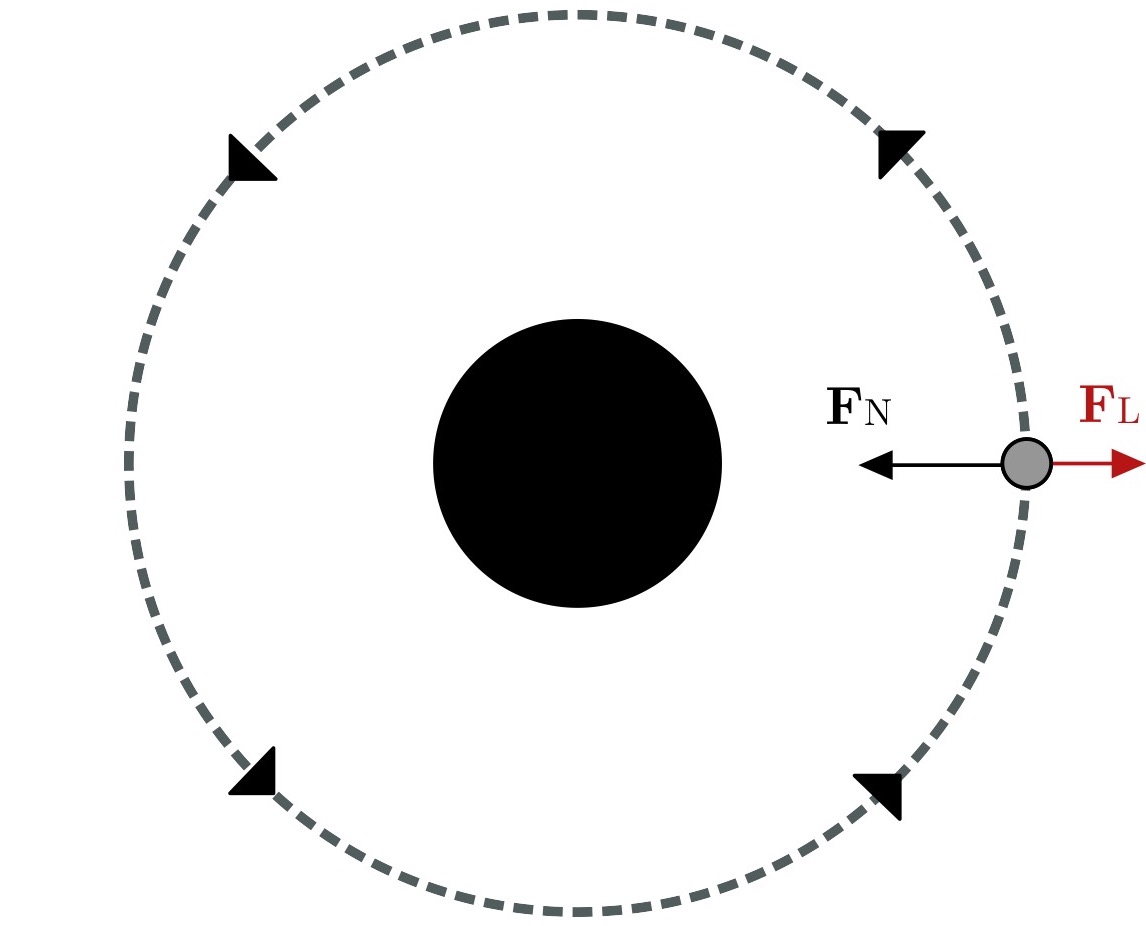}
         \caption{Plus configuration (PC): $\oc > 0$.}
         \label{fig:plus}
     \end{subfigure}
     \hspace{0.07 \textwidth}
   	\caption{Schematic representation of the two configurations of circular orbits discussed in this work. They lie in the equatorial plane of a BH immersed in an asymptotically uniform magnetic field orthogonal to the plane of this page. The positively-charged particle (grey dot) is assumed to orbit the BH (black disk at the center) anticlockwise; the cyclotron frequency is $\oc = q B_0 / m$, where $B_0$ is the magnetic field strength at infinity, $q$ is the charge of the particle, and $m$ is its mass. The configurations are invariant under inversion of the direction of the orbit and of the sign of $\oc$. We also include a Newtonian interpretation of the difference between the two configurations: $\vec{F}_\text{N}$ and $\vec{F}_\text{L}$ represent the gravitational and Lorentz forces, respectively, with the latter being centripetal (centrifugal) in MC (PC) orbits.}
    \label{fig:configurations}
\end{figure*}
In this paper we focus on the circular motion of charged particles in the equatorial plane of a weakly magnetized Schwarzschild BH. The magnetic field is treated as a test field throughout the calculation, inducing no backreaction on the geometry; thus, using standard Schwarzschild coordinates $\{t,r,\theta,\phi \}$, we can write the line element as 
\begin{equation}
ds^2 = - f dt^2 + f^{-1} dr^2 + r^2 d\theta^2 + r^2 \sin^2 \theta \; d\phi^2 \, , 
\label{eq:sch_metric}
\end{equation}
\begin{equation}
f=f(r) = 1 - \frac{2 M}{r} \,, 
\end{equation}
where $M$ is the BH mass. The Schwarzschild solution admits the Killing vector fields 
\begin{equation}
	X^\mu = \delta ^\mu _t \qquad \text{and} \qquad Y^\mu = \delta ^\mu _ \phi \, , 
	\label{eq:Killing_vectors}
\end{equation}
where $\delta^\mu _\nu$ is the identity operator. These vectors are associated with conservation laws for energy and angular momentum along $\hat{z}$, which we define below. In the background above, a stationary and asymptotically uniform magnetic field along the $\hat{z}$ direction corresponds to the vector potential~\cite{PhysRevD.10.1680}
\begin{equation}
A\submu = \frac{B_0}{2} Y\submu = \frac{B_0}{2} r^2 \sin^2 \theta \delta \submu ^\phi\,,\label{eq:sch_mag_potential}
\end{equation}
where $Y^\mu$ is the axial Killing vector defined in Eq.~\eqref{eq:Killing_vectors} and $\vec{B} = B_0 \hat{z}$ is the asymptotic magnetic field. The test field approximation implies that such a solution can only be taken over a region of size~\eqref{eq:limit_validity}.

Consider now a test particle of mass $m$, carrying an electric charge $q$ moving in this spacetime. Following Ref.~\cite{Kolos:2015iva}, we can write the Hamiltonian as 
\be
H \left(\pi\supmu ,  x\supmu\right) =  \frac{1}{2} f \pi_r ^2 + \frac{1}{2r^2} \pi_\theta ^2 - \frac{1}{2} \frac{m^2}{f} \left( \mE^2 - V_{\text{eff}} \right)\, , \label{eq:sch_charged_hamiltonian}
\ee
where $x^\mu=\{t,r,\theta,\phi\}$ are the Schwarzschild coordinates and $\pi^\mu= m u^\mu + q A_\mu$ are the conjugate momenta, with $u^\mu=dx^\mu / d\tau$ the particle's 4-velocity and $\tau$ the proper time. Moreover, we define $\mE$ and $\mL$, the energy and angular momentum along $\hat{z}$ per unit mass,
\begin{equation}
	\begin{split}
		&\mE = \frac{E}{m} = - \frac{\pi_t}{m} = f u^t,  \\ 
		&\mL = \frac{L}{m} = \frac{\pi_\phi}{m} = r^2 \sin^2 \theta \left( u^\phi + \frac{\oc}{2}  \right)
	\end{split}
	\label{eq:sch_conserved_qtys}
\end{equation}
(both conserved quantities), and the effective potential
\begin{equation}
	V_{\text{eff}} = f \left[ 1 + \left( \frac{\mL}{r \sin \theta} - r \sin \theta  \, \frac{\oc}{2} \right)^2 \right]. 
 \label{eq:Veff}
\end{equation}
Note that throughout, instead of the magnetic field strength, we use instead the (possibly negative) cyclotron frequency of the particle, 
\begin{equation}
	\omega_c \equiv \frac{q B_0}{m} \, .
	\label{eq:oc}
\end{equation}
In our numerical analysis, we will be concerned with ratios of fluxes, so results depend only on the cyclotron frequency $\oc$\footnote{Naturally, if we want to actually evolve the orbit of a particle, the three parameters $q$, $B_0$ and $m$ all have to be specified}. For astrophysical magnetic fields~\cite{2011AstBu..66..320P,Eatough:2013nva,Baczko:2016opl,Daly:2019srb}, we find that the cyclotron frequency can grow as large as $M \omega_c \sim 10^{11}$ for an electron orbiting a supermassive BH. We cannot reach this order of magnitude in our simulations, and we only studied values of cyclotron frequency up to $M\oc \sim 10$, which already provides a good understanding of the asymptotic regime of large $\omega_c$. Furthermore, our analytical results are valid for arbitrary cyclotron frequency.

The effective potential governs the region of phase space accessible to the particle, as well as the properties of circular orbits. Circular orbits only exist in the equatorial plane ($\theta=\pi/2$), and we can characterize them via the orbital radius $r_0$, the orbital frequency $\Omega_0$, and the orbital velocity $v_0$. The symmetry of the effective potential, which is invariant under the transformation $(\mL, \oc) \to (-\mL, -\oc)$, motivates the distinction of two different types of circular motion: 

\noindent \emph{Minus Configuration} (MC), where $\mL>0$ and $\oc< 0$. This configuration exists in flat space, as the Lorentz force is centripetal here.

\noindent \emph{Plus Configuration} (PC), where $\mL>0$ and $\oc> 0$. This configuration has no flat-space counterpart, as the Lorentz force is centrifugal (see Fig.~\ref{fig:configurations}).

We choose to always consider positive orbital frequencies and angular momenta $\Omega_0 \,,\mL >0$, so the sign of the cyclotron frequency distinguishes minus ($\oc<0$) and plus ($\oc>0$) configuration orbits. Thus, the orbital frequency can be written as
\begin{align} 
	\Omega_0 
	&= \Bigg[\frac{1}{2+2 r_0^2\omega_c^2} \Big( 2 \omega_K^2 + \omega_c^2 \left(1-2M / r_0 \right) \label{eq:omega_0} \\
	&- \omega_c \sqrt{\omega_c^2\left(1-2M / r_0\right)^2+4 \omega_K^2 \lr{1-3M / r_0}} \Big) \Bigg]^{1/2} \, , \nonumber
\end{align}
where we expressed our result in terms of the cyclotron frequency $\oc$, Eq.~\eqref{eq:oc}, and the Keplerian frequency 
\begin{equation}
\omega_\text{K} = \sqrt{\frac{M}{r_0 ^3}}  \, .\label{eq:ok}
\end{equation}
There are two competing forces, each characterized by a typical frequency: the gravitational force, characterized by $\ok$, and the Lorentz force, characterized by $\oc$. For a given circular orbit, the dominant effect should be associated with the higher frequency; we expect there to be a change of behavior around $\ok\sim\oc$, that is, around a critical radius 
\begin{equation}
	r_c = \lr{\frac{M}{\oc^2}}^{1/3} \, . \label{eq:critical_radius}
\end{equation}

The asymptotic behavior of the orbital frequency
is as follows: for small orbital radius, $r_0 \ll r_c$, it approaches the Keplerian frequency, $\Omega_0 \sim \omega_k$; this limit is expected, since gravitational effects are dominant in this regime. For large orbital radius, $r_0 \gg r_c$, the asymptotic behavior of the orbital frequency for PC orbits is different from that of MC orbits. 
For PC orbits, where the Lorentz force points outwards, we find 
\begin{equation}
\Omega_0 \sim \frac{\omega_\text{K} ^2}{\omega_\text{c}} = \frac{M}{ r_0 ^3 \, \oc} \, .
\label{eq:omega_PC}
\end{equation}
For MC orbits, the Lorentz force points inwards, and so for orbital radius in the range $r_c \ll r_0 \ll 1/M \oc^2$, we recover the standard flat space cyclotron motion~\cite{Jackson:1998nia},
\be
\Omega_0\approx \Omega_\text{c} = - \frac{\omega_\text{c}}{\sqrt{1+(r_0 \omega_\text{c})^2}}\,.\label{eq:omega_MC}
\ee
As $r_0$ increases even further, however, curved space effects become dominant: in fact, expanding $r_0^2\Omega_0^2$ in powers of $1/r_0$ gives
\begin{equation}
r_0^2 \Omega_0^2 \sim 1 - \frac{2M}{r_0} - \frac{1}{r_0^2\omega_c^2} \, , \label{eq:omega_MC_2}
\end{equation} 
as opposed to the flat space expansion $r_0^2 \Omega_c^2 \sim 1 - 1/r_0^2\omega_c^2$. Consequently, for $r_0 \gtrsim 1/M\omega_c^2$ the flat space cyclotron frequency \eqref{eq:omega_MC} stops being a good approximation\footnote{The scale set by $1/M\omega_c^2$ is much larger than the scale set by $r_c$ if we assume that $M\omega_c \ll 1$, i.e., that the characteristic velocity of cyclotron motion at the horizon radius is much smaller than the speed of light.}, and we must use formula~\eqref{eq:omega_MC_2} instead. This happens for ultra-relativistic MC orbits, and can be seen as a manifestation of the redshift effect, since we are using the Schwarzschild time coordinate to define $\Omega_0$. 

\begin{figure*}[t]
     \begin{subfigure}{0.48\textwidth}
         \centering
   		\includegraphics[width= \textwidth]{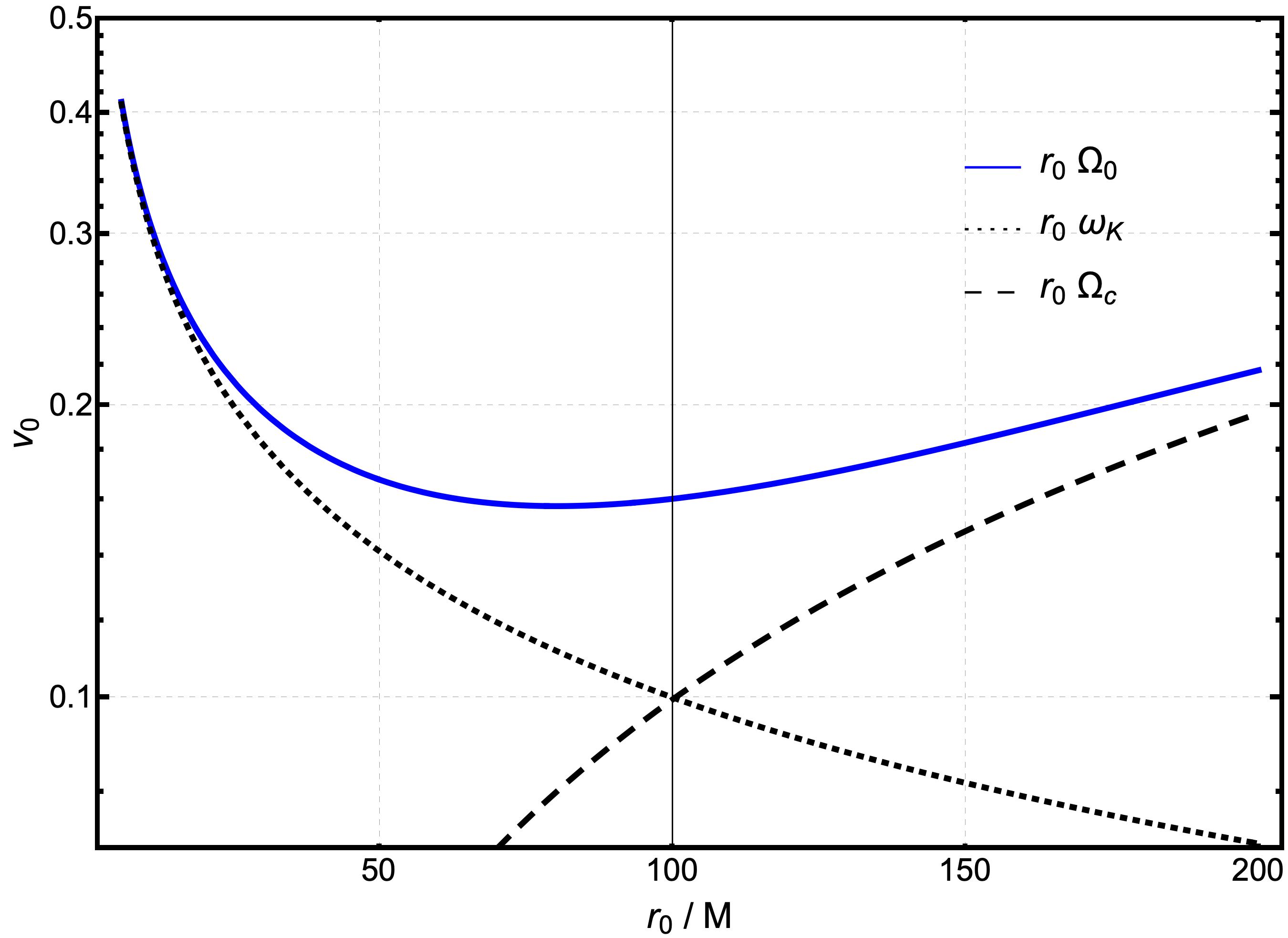}
         \label{fig:vel_prof_3}
     \end{subfigure}
  	\hfill
  	\begin{subfigure}{0.48\textwidth}
         \centering
   		\includegraphics[width= \textwidth]{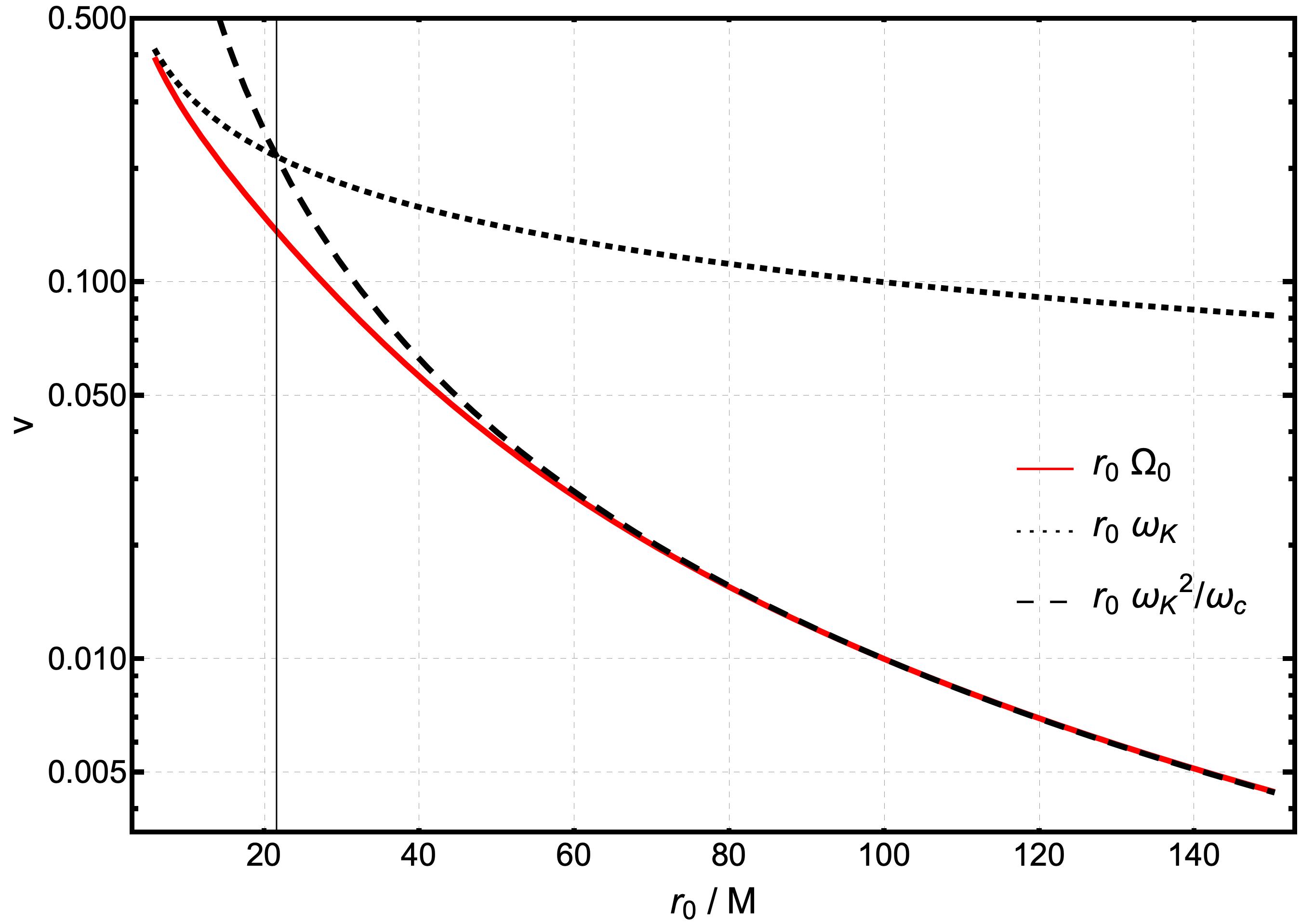}
         \label{fig:vel_prof_4}
     \end{subfigure}
   	\caption{Velocity profiles, as measured by a stationary observer at infinity, of circular orbits with radius $r_0$. Blue and red curves correspond to the velocity profiles for MC and PC orbits, respectively.
{\bf Left panel:} MC orbits for $M |\oc| = 10^{-3}$. 
{\bf Right panel:} PC orbits for $M |\oc| = 10^{-2}$. The solid blue and red curves are calculated using Eq.~\eqref{eq:omega_0}; the black dashed curves are obtained by taking the orbital frequency to be the Keplerian frequency (Eq.~\eqref{eq:ok}); the black dashed curves are obtained using Eq.~\eqref{eq:omega_MC} in MC orbits and Eq.~\eqref{eq:omega_PC} in PC orbits. Finally, the black solid thin vertical line indicates the critical radius (Eq.~\eqref{eq:critical_radius}). We show different values of $|M \oc|$ in the two panels because these appropriate to study each configuration and are thus used further ahead in Sec.~\ref{sec:numerical}. In the right panel, the Keplerian frequency is only a good approximation very close to the ISCO because $|M \oc|$ is large enough to change the velocity profile for smaller orbital radius than in the left pannel (as seen by the position of $r_c$).} 
    \label{fig:vel_prof}
\end{figure*}
Finally, all orbits are stable up to the innermost stable circular orbit (ISCO). We find that taking the limit $|M \oc| \to \infty$ leads to 
\beq
&&r_0^\text{ISCO} \to 2M\,,\,\,\qquad \qquad \qquad \text{PC orbits,}\\
&&r_0 ^\text{ISCO}  \to \frac{\lr{5+\sqrt{13}}M}{2} \,,\qquad \text{MC orbits.}
\eeq
Regarding the orbital velocity $v_0=r_0 \Omega_0$ of the particles at the ISCO, as measured by stationary observers at infinity, it tends to zero for PC orbits (essentially due to gravitational time dilation), and to a finite value for MC orbits, where we get
\begin{equation}
v_0 ^\text{ISCO} \to \frac{\sqrt{6+\lr{3+\sqrt{13}}}}{5+\sqrt{13}} \approx 0.732 \qquad (|M \oc|\to \infty) \, .
\end{equation} 
These results are compatible with those presented in Refs.~\cite{Aliev:1980hz,Frolov:2010mi,Frolov:2011ea,Zahrani:2013up,Kolos:2015iva,Qi:2023brf,Baker:2023gdc}.

For completeness, we also look at the velocity profiles in MC and PC orbits (see Fig.~\ref{fig:vel_prof}). 
Changing $\oc$ implies a shift in $r_c$ and a subsequent shift of the velocity profile. In MC orbits, the velocity increases for large radius and asymptotic orbits are relativistic; this happens because the Lorentz force is attractive and dominant in this region and so, just like in flat space cyclotron motion, the particle must move faster in wider orbits. In PC orbits, the velocity decreases for all orbital radii; the Lorentz force is now repulsive so it decreases the centripetal force.

It is easy to show from Eqs.~\eqref{eq:sch_conserved_qtys}--\eqref{eq:omega_0} that circular orbits with larger radius have larger energy. Thus, if the particle is in a circular orbit and radiates positive energy, then its energy decreases and the orbital radius must decrease. Conversely, if the particle gains energy, the orbital radius must increase.

\subsection{Radiation reaction force in curved space}
The generalization to curved space of the dynamics of a radiating charged particle of mass $m$ and charge $q$ in the presence of an EM field yields the DeWitt-Brehme equation~\cite{DEWITT1960220, HOBBS1968141,Poisson:2011nh}. This is a third order differential equation, and so a reduction of order must be performed to avoid runaway solutions~\cite{Flanagan:1996gw,Spohn:1999uf,2001PhLA..283..276R,Gralla:2009md}; doing so, and taking the metric to be a vacuum solution of the Einstein equations ($R\submunu = 0$), yields
\beq
&&\frac{D u\supmu}{d\tau} = \frac{q}{m} F\supmu _{\ \ \nu} u\supnu + \frac{2 q^2}{3 m} \left(\frac{q}{m} \nabla_\alpha F^\mu _{\ \ \nu} u^\alpha u^\nu\right) \label{eq:rad_motion_reduced} \\
&+& \frac{2 q^2}{3 m} \left(\frac{q^2}{m^2} \left(F^{\mu \nu}F_{\nu \rho} + F^{\nu \alpha}F_{\alpha \rho} u_\nu u^\mu \right)u^\rho \right) + \frac{2 q^2}{ m}f^{\mu \nu}_\text{tail} u_\nu \,. \nonumber
\eeq
Here, the only difference with respect to the reduced Abraham-Lorentz-Dirac equation in flat space is the tail term, 
\begin{equation}
	f^{\mu \nu}_\text{tail} = \int _{-\infty} ^{\tau^-} \nabla^{[ \mu} G^{\nu]}_{\ \ \lambda^\prime} \lr{z(\tau), z(\tau^\prime)} u^{\lambda^\prime} d \tau^\prime \, ,
	\label{eq:rad_tail}
\end{equation}
where square brackets denote anti-symmetrization, $z(\tau)$ is the particle's  worldline, $G^{\nu}_{\ \ \lambda^\prime} \lr{x,x^\prime}$ is the retarded Green's function for the vector wave equation in curved space, and the integral is taken over the entire history of the particle. A modern derivation of this formula is given by Poisson~\cite{Poisson:2011nh}.

The non-local tail term is very complicated to calculate~\cite{DeWitt:1964de, Smith:1980tv}. Previous analysis estimated its magnitude based on a static-charge approximation, and concluded that it can be neglected~\cite{2018ApJ...861....2T}. Unfortunately, we are precisely interested in moving charges where such an argument does not hold~\cite{Santos:2023uka}.

%
%

To make the point clear, let us assume that the tail term can be neglected.
The inclusion of radiation force in Eq.~\eqref{eq:rad_motion_reduced} without the tail term causes the particle energy to change with time according to 
\begin{equation}
\frac{d E}{d t} = -\frac{2q^2}{3} \oc f \left[ \oc \mE^2 - \left(\oc f + \frac{M}{r}u^\phi \right)  \right] \,,\label{eq:sch_energy_change}
\end{equation}
which allows the particle to {\em gain} energy, that is, $dE/dt>0$. Indeed, for circular orbits of radius $r_0$, the condition $dE/dt>0$ can be written as
\begin{equation}
\oc >0 \quad \text{and} \quad 0 < f(r_0)  u^\phi \ < \ok^2 / \oc \, .\label{eq:sch_energy_gain_PC}
\end{equation}
This condition is satisfied for all PC orbits outside the light ring ($r_0 > 3M$). Since the energy of circular orbits increases with the radius of the orbit, we expect the orbital radius of a particle in a PC orbit to increase as it gains energy according to Eq.~\eqref{eq:sch_energy_change}. This ``orbital widening" was reported in Refs.~\cite{2018ApJ...861....2T,Tursunov:2018udx}.

This result is counter-intuitive (and wrong!): there is no energy conservation, since there is no source of energy in the system that could be feeding the ``orbital widening". In the upcoming sections we will show that ``orbital widening" does not take place. We will also show that the tail {\it can} be neglect in some circumstances, but never in PC orbits, hence ``orbital widening" is not possible. We will employ BH perturbation theory~\cite{Teukolsky:1972my,Teukolsky:1973ha,Bardeen:1973xb}. This will allow us to calculate the radiation field of a particle in circular orbit and the corresponding energy flux at infinity and on the BH horizon, which will be compared with Eq.~\eqref{eq:sch_energy_change}, allowing us to evolve the circular orbit in the adiabatic approximation. Notice that this formalism is fully relativistic, and so the results are valid in the strong field regime.


\section{Electromagnetic fields on a Schwarzschild geometry} \label{sec:perturbations}
%
\subsection{The Teukolsky equation}
The Teukolsky equation~\cite{Teukolsky:1972my} is derived within the Newman-Penrose formalism~\cite{doi:10.1063/1.1724257}, and describes the radiative degrees of freedom of massless fields of spin-weight $s$ ($s=0,-1,-2$ for scalars, vectors and tensors, respectively) in the background of a rotating Kerr BH. Setting the BH angular momentum to zero yields the Bardeen-Press equation~\cite{Bardeen:1973xb}, which in Schwarzschild coordinates can be written as 
\beq
&&	\lr{\frac{r^2}{f}} \partial_t ^2 \psi - \frac{1}{\sin^2 \theta} \partial _\phi ^2 \psi 
-\lr{\frac{1}{r^2 f}}^s \partial_r \lr{\lr{r^2 f}^{s+1}\partial_r \psi}\nonumber \\
&&  -\frac{1}{\sin \theta} \partial_\theta \lr{\sin \theta \ \partial_\theta \psi} - 2 i s \lr{\frac{\cos \theta}{\sin^2 \theta}} \partial_\phi \psi \label{eq:teukolsky} \\
&& - 2 s \lr{\frac{M}{f}-r} \partial_t \psi  + \lr{s^2 \cot^2 \theta -s}\psi = 4 \pi r^2 T \, , \nonumber
\eeq
%
%
%
%
where $\psi$ is the perturbation field and $T$ is a source term. The $t$ and $\phi$ dependencies are trivial, while the radial and angular parts are separable if the fields are expanded in terms of spin-weighted spherical harmonics $_sY_{\ell m}(\theta,\phi)$~\cite{1967JMP.....8.2155G}. Therefore, we start by expanding the fields and sources in their Fourier-harmonic components 
\begin{equation}
	\psi = \int _{- \infty} ^\infty d \omega \sum _{\ell,m } ^\infty R_{\omega \ell m } (r) _s Y_{\ell m} (\theta, \phi) e^{-i \omega t} \, ,
	\label{eq:psi_fourier_harmonic}
\end{equation}
\begin{equation}
	T = \int _{- \infty} ^\infty d \omega \sum _{\ell,m } ^\infty T_{\omega \ell m } (r) _s Y_{\ell m} (\theta, \phi) e^{-i \omega t} \, ,
	\label{eq:T_fourier_harmonic}
\end{equation}
and obtain the equation for the radial functions $R_{\omega \ell  m}$ (we will drop the $\omega \ell m$ subscripts from now onwards, for simplicity):
\beq
&&\lr{r^2 f \frac{d^2}{d r^2} + 2 (s+1)(r-M)\frac{d}{dr} - V_s}R = 4 \pi r^2 T\,,\label{eq:sch_teukolsky_radial}\\
&&V_s(r) = (\ell-s)(\ell+s+1)- \frac{(r \omega)^2}{f} + 2 i \omega s \lr{\frac{M}{f} - r} \, . \nonumber
\eeq
This equation can be solved by taking two linearly independent solutions of the homogeneous radial equation~\eqref{eq:sch_teukolsky_radial}. We label these solutions $R^H$ and $R^\infty$ because we will see that they correspond to the boundary conditions on the horizon and at infinity, respectively. From these two solutions of the homogeneous equations we can define a rescaled Wronskian
\begin{equation}
\mathcal{W}=(r^2 f)^{s+1} \lr{R^\infty \frac{d R^H 
}{d r}-R^H \frac{d R^\infty}{d r}}\, , \label{eq:Wronskian}
\end{equation}
which is constant. We can then find the solution of the radial equation with a source with the appropriate asymptotic behavior on the horizon and at infinity in the form
\beq
R(r) &=& \frac{R ^\infty (r)}{\mathcal{W}}  \int_{2 M} ^r R ^H (r^\prime) T(r^\prime) \ (r'^2 f(r^\prime))^s d r^\prime \nonumber\\
&+& \frac{R ^H (r)}{\mathcal{W}} \int_{r} ^\infty R ^\infty (r^\prime) T (r^\prime) \ (r'^2 f(r^\prime))^s d r^\prime \, . \label{eq:gen_sol_r}
\eeq
%
\subsection{Electromagnetic perturbations of Schwarzschild spacetime} \label{sec:EM_perturbations}
%
We are interested in studying EM perturbations on a Schwarzschild background. To use the Teukolsky equation~\eqref{eq:teukolsky} we must first decompose the EM field into components with a well-defined spin weight $s$. It turns out that all the relevant quantities (energy flux at infinity and on the horizon) can be expressed in terms of a single component, so that in the end we just have to solve Eq.\eqref{eq:teukolsky} with the substitutions 
\begin{align}
\psi = r^2 \phi_2 \, , \quad s = -1  \, , \quad 
T = r^2 J_2\,, \label{eq:sub_teuk}
\end{align}
where $\phi_2$ and $J_2$ are $s=-1$ quantities built from the Faraday tensor and the charge 4-current, respectively \cite{Teukolsky:1973ha}. 

On physical grounds, the radiation field is expected to correspond to purely outgoing waves at infinity and purely ingoing waves on the horizon. This must be imposed at the level of the solutions of the homogeneous radial equation $R^H$ and $R^\infty$, as seen in Eq.~\eqref{eq:gen_sol_r}. The asymptotic behavior of the solutions of the homogeneous radial equation for $s=-1$ is \cite{Teukolsky:1973ha} 
\beq
R^H &\sim & \frac{A_{\text{in}}}{r} \ e^{- i \omega r_\star}+ A_{\text{out}}\ r \ e^{ i \omega r_\star} \, , \nonumber \\
R^\infty &\sim&  r \ e^{i \omega r_\star}\,,\qquad r\to\infty \label{eq:phi2_homo_ass_infty} 
\eeq
and
\beq
R^\infty &\sim& B_{\text{in}} r^2 f e^{- i \omega r_\star} + B_{\text{out}} e^{ i \omega r_\star} \, ,  \nonumber \\
R^H & \sim &r^2 f \ e^{- i \omega r_\star}\,,\qquad r\to 2M\,, \label{eq:phi2_homo_ass_2M}
\eeq
where $r_\star$ is the usual tortoise coordinate. Next, we must take care of the source term. 
%
\subsection{Energy radiated by a charged particle in a circular orbit} \label{sec:EM_energy_circular_orbit}
%
The 4-current $J_\mu$ of a single particle of charge $q$ is
\begin{equation}
J_\mu (x) = q \int d \tau \ u_\mu (\tau) \delta^{(4)}(x-z(\tau)) \, ,\label{eq:Jmu}
\end{equation}
where $z(\tau)$ is the particle's trajectory, $\tau$ is the proper time and $u^\mu = dz^\mu /d\tau$ is the 4-velocity. In our particular case, these should correspond to a particle in a circular orbit in the equatorial plane with radius $r_0$ and orbital velocity $\Omega_0 = d \phi / dt$. We then find
\begin{equation}
	J_\mu (x) = q \frac{u_\mu}{u^t} \frac{\delta (r-r_0)}{r^2} \frac{\delta(\cos \theta)}{\sin \theta} \delta(\phi - \Omega_0 t)\, .
\end{equation}
This result must be replaced into the expression for the source term $T=r^2 J_2$, which we then Fourier-expand as indicated in Eq.~\eqref{eq:T_fourier_harmonic}. One gets
\begin{widetext}
\begin{align}
&\phi_2 = \frac{1}{r^2} \sum_{\ell, m} \left[R^\infty (r) \, Z^\infty \, \Theta(r-r_0) +  R^H(r) \, Z^H \,\Theta(r_0-r)\right] Y_{\ell m} (\theta,\phi) \, e^{- i m\Omega_0 t} \, , \label{eq:phi2} \\
&Z^{\infty,H}\equiv   \frac{i \pi q}{\sqrt{2} \, m \Omega_0 A_{\text{in}}} \Bigg[ \frac{R^{H,\infty}(r_0)}{ f(r_0)} \Bigg(\bigg( i m \Omega_0 + \frac{3 f(r_0)}{r_0} \bigg) i v_0 \ _{-1}\overbar{Y}_{\ell m}(\pi/2,0) + \frac{\sqrt{\ell(\ell +1)}}{ r_0} f(r_0) \ _0  \overbar{Y}_{\ell m}(\pi/2,0) \Bigg) \nonumber \\
& \qquad \qquad \qquad \qquad \qquad - i v_0 \frac{1}{r^2} \frac{d}{d r}\big(r^2 R^{H,\infty} \big)_{r=r_0} \  _{-1}\overbar{Y}_{\ell m}(\pi/2,0)\Bigg] \, ,\label{eq:sch_teukolsky_Zlm_inf}	
\end{align}
\end{widetext}
where $\Theta(x)$ is the Heaviside step function, and the frequency is $\omega=m\Omega_0$ (we remind the reader that the homogeneous functions $R^{H,\infty}$ carry $\omega \ell m$ subscripts). 

Once the field is determined, we can use it to compute energy fluxes~\cite{Teukolsky:1974yv},
\begin{equation}
	\dot{E}^\infty  \equiv \frac{d E}{dt}\Bigg|_{\infty} = \frac{1}{2 \pi} \sum_{\ell ,m} \left| Z^\infty _{\ell m} \right|^2
	\label{eq:sch_teukolsky_phi2_ene_flux_inf}
\end{equation}
and
\begin{equation}
	\dot{E}^H = \sum_{\ell ,m} \frac{32 \, (m\Omega_0)^2 M^6  (16 \,(m \Omega_0)^2 + 1/M^2)}{\pi \left[ \ell(\ell +1)\right]^2}  \left| Z^H  _{\ell m} \right| ^2 \, ,
	\label{eq:sch_teukolsky_phi2_ene_flux_hor}
\end{equation}
respectively. Despite their complicated appearance, the only difficult task is actually finding the solutions $R^{H,\infty}$ of the homogeneous radial equation, Eq.~\eqref{eq:sch_teukolsky_radial}, with physical boundary conditions. In the following we will employ both analytical and numerical methods to accomplish this task, and to obtain the energy fluxes. 

We recall that we calculate these energy fluxes for a particle in a circular orbit with the aim of studying particle motion using the adiabatic approximation, that is, assuming the particle is always in a circular orbit. However, for this to be true, energy and angular momentum losses must be such that they continue driving the particle on a circular path. It turns out, the energy and angular momentum carried out by a mode with azimuthal number $m$ and frequency $\omega$ are related by~\cite{Bekenstein:1973mi, Teukolsky:1974yv,Cardoso:2020iji} 
\begin{equation}
\dot{L} = \frac{m}{\omega} \dot{E}\, .\label{eq:flat_momentum_change}
\end{equation}
In our case we have $\omega = m \Omega_0$ and we find indeed that the particle will remain in (quasi-)circular motion upon adiabatic backreaction from electromagnetic radiation~\cite{Cardoso:2020iji}.
%
\section{Analytical results} \label{sec:analytical}
%
We have established that a particle in a circular orbit
around a Schwarzschild BH loses energy through radiation to infinity and into the BH, given by Eqs.~\eqref{eq:sch_teukolsky_Zlm_inf}-- \eqref{eq:sch_teukolsky_phi2_ene_flux_hor}. The problem of actually finding the value of the energy flux for a given set of parameters really boils down to finding $R^H (r_0)$, $R^\infty (r_0)$ and $A_\text{in}$, that is, it reduces to solving the homogeneous Teukolsky equation with physical boundary conditions. This can be done analytically in the small frequency limit, $M \omega \ll 1$, for slow motion, $v_0 = r_0 \Omega_0 \ll 1$. The corresponding calculations are presented in Appendix~\ref{app:matched_asymptotic_expansions}; here we simply use these results to compute the energy fluxes.
%
\subsection{Energy flux at infinity} \label{sec:anres_infinity}
Consider first the low-frequency, closed-form solution for the energy flux at infinity from the dipole mode $\ell=|m|=1$ term, obtained from Eqs.~\eqref{eq:sch_teukolsky_Zlm_inf} and~\eqref{eq:sch_teukolsky_phi2_ene_flux_inf}. This is the dominant contribution at low velocities, and yields a generalized Larmor formula (GLF):
\begin{equation}
\dot{E}^\infty_{11}+\dot{E}^\infty_{1-1} = \frac{1}{\pi} \left| Z^\infty _{11} \right|^2 \approx \frac{2}{3} q^2 (r_0-2 M)^2 \Omega_0^4 \, .\label{eq:anres_ene_flux_inf}
\end{equation}
Note that $\dot{E}^\infty_{\ell m}=\dot{E}^\infty_{\ell-m}$~\cite{Cardoso:2019nis}. We call \eqref{eq:anres_ene_flux_inf} a GLF because it reduces to the Larmor formula for large $r_0$, the only difference between them being the scaling $\propto (r_0 -2)^2$, which implies the energy flux at infinity goes to zero as the orbit approaches the BH event horizon.

We can also obtain expressions for the energy flux at infinity in a generic mode $(\ell, m)$. It turns out we must treat modes with even and odd values of $\ell+m$ separately~\cite{Poisson:1994yf}. This is because the dominant term in the low-frequency and slow orbit approximation is different for the two cases. We find
\begin{widetext}
\beq
\dot{E}^\infty _{\ell m}&=& 2^{4 \ell-4} \frac{q^2}{M^2} (r_0/M-2 )^2  (m M \Omega_0)^{2(\ell +1)} \frac{ (\ell +1) \, (2 \ell +1) \, \Gamma (l)^2 \, \Gamma (\ell +1)^2 \, \Gamma (\ell +2)^2 \, (\ell-m)! \, (\ell +m)! }{\ell \, \Gamma (2 \ell)^2 \, \Gamma (2 \ell +2)^2 \, ((\ell-m)\text{!!})^2 \, ((\ell +m)\text{!!})^2} \nonumber\\
& \times& \ _2F_1\left(1-\ell,\ell +2;2;1-\frac{r_0}{2 M}\right)^2 \, ,\label{eq:anres_ene_flux_inf_even}
\eeq
\beq
\dot{E}^\infty _{\ell m}&= & 2^{4 \ell-8} \frac{q^2}{M^2} m^2 v_0^4 (m M \Omega_0)^{2\ell}  \ \frac{ (2 \ell +1) \, \Gamma (\ell)^2 \, \Gamma (\ell +1)^2 \, \Gamma (\ell +2)^2 \, ((\ell - m)\text{!!})^2 \, ((\ell +m)\text{!!})^2 }{\ell^3 \, (\ell +1) \, \Gamma (2 \ell)^2 \, \Gamma (2 \ell +2)^2 \,(\ell-m)! \, (\ell +m)!} \nonumber \\
&\times& \left(\left(\ell^2+\ell-2\right)  (r_0/M -2 )  \, _2F_1\left(2-\ell,\ell +3;3;1-\frac{r_0}{2 M}\right) +4   \, _2F_1\left(1-\ell,\ell +2;2;1-\frac{r_0}{2 M}\right)\right)^2 \label{eq:anres_ene_flux_inf_odd}
\eeq
\end{widetext}
for even and odd values of $\ell + m $, respectively, where $F$ is a standard hypergeometric function of the second kind.

We have performed a detailed comparison of the relative importance of different multipolar modes and concluded that, within the domain of validity of the slow-motion and low-frequency approximation, the dipole term dominates the energy flux at infinity, even in the strong field regime, up to about $r_0 \approx 2.1 $. Beyond that value, higher multipoles must be included to get an accurate estimate of the energy flux at infinity. This dominance is very clear if we take $r_0\to \infty$ and compare the flux in a given mode to the dipolar flux,
\begin{widetext}
\begin{equation}
\frac{\dot{E}^\infty _{\ell m}}{\dot{E}^\infty _{1 1}} \sim 3\times 4^\ell  m^{2 \ell +2} \, \frac{ \ell \, (\ell +1) \, (2 \ell +1) \, \Gamma (\ell)^2 \, \Gamma (\ell - m+1)\,  \Gamma (\ell +m+1) }{((\ell - m)\text{!!})^2   \, ((\ell +m)\text{!!})^2 \, \Gamma (2 \ell +2)^2 } (v_0)^{2 \ell-2}\,,    \qquad r_0 \to \infty \label{eq:anres_ene_flux_inf_even_inf} \, ,
\end{equation}
\begin{equation}
\frac{\dot{E}^\infty _{\ell m}}{\dot{E}_{1 1} ^\infty}  \sim  3\times 4^{\ell +1} m^{2 \ell +2} \frac{ \ell (\ell +1) (2 \ell +1) \Gamma (\ell)^2 ((\ell - m)\text{!!})^2  ((\ell +m)\text{!!})^2 }{\Gamma (2 \ell +3)^2 \, \Gamma (\ell - m+1) \, \Gamma (\ell +m+1)} (v_0)^{2 \ell} \,, \qquad r_0 \to \infty
   \label{eq:anres_ene_flux_inf_odd_inf} \, , 
\end{equation}
\end{widetext}
for even and odd values of $\ell + m $, respectively. The scaling with $v_0$ coincides with that for gravitational radiation~\cite{Poisson:1994yf}. 
%
%
\subsection{Energy flux across the horizon} \label{sec:anres_horizon}
%
We now turn to the energy flux across the BH horizon, Eqs.~\eqref{eq:sch_teukolsky_Zlm_inf}--\eqref{eq:sch_teukolsky_phi2_ene_flux_hor}. We find the general (small-frequency, small orbital velocity) result
\begin{widetext}
\beq
\dot{E}^H _{\ell m}&= & 2^{2 \ell +2} \frac{q^2}{M^2} (r_0/M-2 )^{-2 (\ell +1)} (m M \Omega_0)^2 \frac{ (2 \ell +1) \, \Gamma (\ell +1)^2 \, \Gamma (\ell +2)^2 \, (\ell - m)! \, (\ell +m)! \,}{\ell \, (\ell +1) \, \Gamma (2 \ell +2)^2 \, ((\ell - m)\text{!!})^2 \, ((\ell +m)\text{!!})^2}\nonumber \\
	& \times& _2F_1\left(\ell +1,\ell +2;2(\ell +1);-\frac{2 M}{r_0-2M}\right)^2  \, , \label{eq:anres_ene_flux_horizon_even}
\eeq
\beq
\dot{E}^H _{\ell m}  &=& 2^{2 \ell +2} \frac{q^2}{M^2} (r_0/M-2)^{-2 (\ell +3)} m^2 v_0^4 \frac{ (2 \ell +1) \Gamma (\ell +1)^2 \Gamma (\ell +2)^2 ((\ell - m)\text{!!})^2 ((\ell +m)\text{!!})^2 }{\ell^3 (\ell +1)^3 \Gamma (2 \ell +2)^2 (\ell - m)!
   (\ell +m)!} \label{eq:anres_ene_flux_horizon_odd} \\
   & \times & \bigg((\ell+1)(r_0/M -2 )
   \, _2F_1\left(\ell +1,\ell +2;2 (\ell +1);-\frac{2 M}{r_0-2M}\right) -(\ell +2) \, _2F_1\left(\ell +2,\ell +3;2 \ell +3;-\frac{2M}{r_0-2M}\right)\bigg)^2 \nonumber
\eeq
\end{widetext}
for even and odd values of $\ell + m$, respectively. Then, it is easy to show that for $r_0 > 2.1M$, the first few modes with $m=\ell$ are enough to capture the radiation absorbed by the BH (provided, of course, that the orbits are slow). On the contrary, very close to the BH ($r_0 < 2.1M$), even if the orbits are slow, we must include more and more modes to obtain the total energy flux on the horizon, eventually leading to $\omega = m \Omega_0 \sim 1$, thus breaking our approximation. Taking $\ell=m=1$ above, that is, looking just at the energy flux in the dipole, mode yields
\begin{equation}
	\dot{E}_{11} ^H  = 3 q^2 \Omega_0^2 \frac{\left(2M(M- r_0)+ r_0(r_0-2 M) \log \left(\frac{r_0}{r_0-2 M}\right)\right)^2}{4 M^2 r_0 ^2} \,.
	\label{eq:anres_ene_flux_horizon}
\end{equation}
Note that now, in contrast to what we discussed for the flux at infinity in Eq.~\eqref{eq:anres_ene_flux_inf}, the flux on the horizon goes to a finite value when $r_0 \to 2 M $. 
%
\subsection{The horizon dominance effect} \label{sec:anres_compare}
%
Hand-waving arguments suggest that fluxes at the horizon are in general much smaller than those at infinity. This is especially true when the objects are far apart, since then each occupies but a small fraction of the other's sky. Surprisingly, this argument fails in some circumstances. 

Consider the ratio of fluxes, when $r_0\to \infty$, 
\begin{align}
&\hspace{-0.7 em} \frac{\dot{E} _{\ell m} ^H}{\dot{E} _{\ell m}^\infty }  \sim 4  \left(\frac{r_0}{M}\right)^{-4 \ell-2} (m M \Omega_0)^{-2 \ell} \left(\ell \, \Gamma (\ell)\right)^2   \label{eq:anres_ene_flux_horizon_even_inf}  \, , \\
&\hspace{-0.7 em} \frac{\dot{E} _{\ell m} ^H}{\dot{E} _{\ell m}^\infty } \sim 4  \left(\frac{r_0}{M}\right)^{-4 \ell-2} (m M \Omega_0)^{-2\ell} \left( (\ell +1)  \Gamma (\ell)\right) ^2 \label{eq:anres_ene_flux_horizon_odd_inf} 
\end{align}
for even and odd values of $\ell +m$, respectively. 
Together with Eqs.~\eqref{eq:anres_ene_flux_inf_even_inf}--\eqref{eq:anres_ene_flux_inf_odd_inf}, these show that, when $r_0 \to \infty$, the dipolar mode dominates energy emission both at infinity and at the BH horizon. For Keplerian orbits, these results coincide with those of Poisson and Sasaki~\cite{Poisson:1994yf}.
In particular, when the particle is in free fall ($\oc=0$) and the angular frequency is $\Omega_0 = \omega_K$, we have
\begin{equation}
\frac{\dot{E}_{1 1}^H }{ \dot{E}_{1 1}^\infty } \sim \frac{4 M^4}{r_0 ^6 \Omega_0^2} \ \xrightarrow[]{\ \Omega_0 = \ok \ } \ \frac{4 M^3}{r_0 ^3}\,, \qquad r_0 \to \infty \, .
	\label{eq:anres_ratio_horizon_inf_B0}
\end{equation}
For this type of orbit, the flux at infinity greatly dominates over horizon fluxes.

However, Eq.~\eqref{eq:anres_ratio_horizon_inf_B0} shows that if the angular frequency decays faster than $1/r_0^3$, then the horizon flux can dominate. The bordering case $\Omega_0 \propto r_0 ^{-3}$ is precisely what we got for charged particles in PC orbits in the region $r_0 \gg r_c$, Eqs.~\eqref{eq:critical_radius}--\eqref{eq:omega_PC}. Thus, for PC orbits,
\begin{equation}
\frac{\dot{E}  _{1 1}^H }{ \dot{E}_{1 1}  ^\infty } \sim \frac{4 M^4}{r_0 ^6 \Omega_0^2} \ \xrightarrow[]{\ \Omega_0 = \ok^2 / \oc \ } \ 4 M^2 \oc ^2\,, \qquad r_0 \to \infty\, .\label{eq:anres_ratio_horizon_inf_PC}
\end{equation}
The ratio asymptotes to a constant as the particle orbits farther and farther away!
The validity of the result above relies on orbits being slow and radiation having a small frequency, both of which hold. For higher order multipoles, Eq.~\eqref{eq:anres_ene_flux_horizon_even_inf} and \eqref{eq:anres_ene_flux_horizon_odd_inf} show that the flux on the horizon always becomes dominant for very wide PC orbits.

The ``horizon dominance effect" we are reporting is not attached specifically to the magnetic field: it is rather a general statement. For a given value orbital radius, by decreasing the velocity one increases the fraction of the total energy absorbed by the BH. In fact, this phenomenon had already been partly noticed in~\cite{Galtsov:1982hwm,Brito:2012gw}.

We show in Appendix~\ref{app:scalar_grav} (which uses results from Appendix\ref{app:matched_asymptotic_expansions}) that these conclusions apply to other massless fields, such as scalar and gravitational radiation. Notice that horizon dominance may occur when the orbiting object is arbitrarily far away from the BH, in a region where spacetime is essentially flat. These results motivate the study of a simpler system sharing the same features. We consider in Appendix~\ref{app:flat_space} an absorbing sphere in flat space, and show that it behaves, essentially, in the same way. Fluxes across its surface can vastly dominate over fluxes at infinity.

%
\subsection{Importance of the tail term} \label{sec:imp_tail}
%
In a previous work we had studied the same system but restricted to the Newtonian limit: weak gravitational and magnetic field and slow motion~\cite{Santos:2023uka}. The results above reduce to those of Ref.~\cite{Santos:2023uka}, in their regime of validity.

In this work, we extended the domain of validity of our previous results. We can now strengthen the claim that the tail term must, in general, be included. In particular, the ``orbital widening" reported in Ref.~\cite{2018ApJ...861....2T} cannot take place. This can be seen, for example, from Eqs.~\eqref{eq:sch_teukolsky_phi2_ene_flux_inf} and~\eqref{eq:sch_teukolsky_phi2_ene_flux_hor}, which are manifestly non negative; thus, energy conservation requires that the particle must be \emph{losing} energy. Since the energy of the particle increases with the orbital radius (see Eq.~\eqref{eq:Veff}), we conclude the orbit must be shrinking. ``Orbital widening" cannot take place. 

%
%
\section{Numerical results} \label{sec:numerical}
%
We now proceed to study the EM radiation using numerical methods, therefore valid for generic frequencies and orbital motion. Our numerical procedure has been reported elsewhere and matches well other results in the literature~\cite{Cardoso:2021vjq,Cardoso:2022fbq,Cardoso:2022whc}.

%
%
\subsection{Minus configuration orbits: $\oc < 0$}
%
%
%
%
\begin{figure}
\centering
\includegraphics[width= .5 \textwidth]{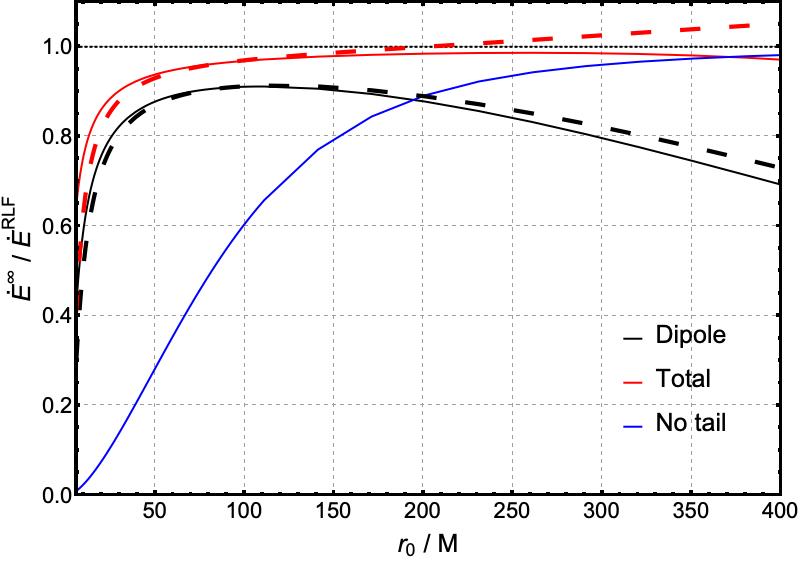}
\caption{Energy flux of EM radiation at infinity normalized to the RLF value~\eqref{eq:rel_larmor_circular}, for a charged particle in an MC circular orbit of radius $r_0$ around a BH with $M \oc=-10^{-3}$. Solid black curve gives dipolar $\ell=1$ flux, solid red curve includes also the sum of the quadrupole $\ell=2$ and octupole $\ell=3$, where we sum the modes with symmetric values of $m$. Dashed lines represent analytical predictions~\eqref{eq:anres_ene_flux_inf_even}, both for the dipole and for the sum up to the octupole. 
Lowest orbital radius $r_0$ corresponds to the ISCO. The sum of modes allows us to recover the RLF with very good accuracy in a wide range of orbits. 
Blue curve is prediction~\eqref{eq:sch_energy_change}, which neglects the tail term. For small $r_0$, one cannot recover the correct results if the tail is neglected.}
\label{fig:numerical_minus_infinity}
\end{figure}
\begin{figure}
    \centering
    \includegraphics[width=.5 \textwidth]{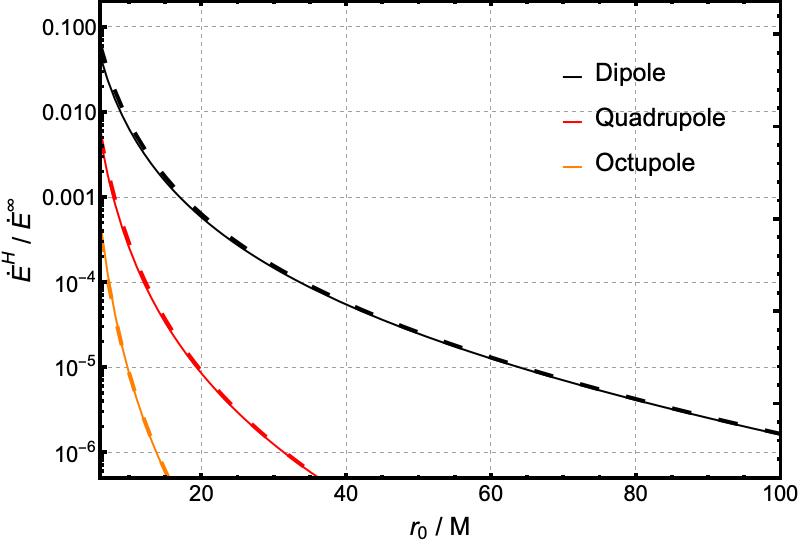}
    \caption{Ratio between the energy flux of EM radiation on the horizon and at infinity, for a charged particle in a MC circular orbit of radius $r_0$ around a Schwarzschild BH with $M \oc=-10^{-3} $. The solid black, red and orange curves correspond to the dipole, quadrupole, and octupole modes. The dashed lines are the predictions for the three modes obtained analytically using~\eqref{eq:anres_ene_flux_horizon_even} and~\eqref{eq:anres_ene_flux_inf_even}. Lowest orbital radius $r_0$ corresponds to the ISCO. We find good agreement between numerical and analytical results. The dipole term is dominant, and we find that the energy flux on the horizon is non-negligible only for $r_0 / M \lesssim 10$.}
    \label{fig:numerical_minus_horizon_ratio}
\end{figure}

Our numerical results are summarized in Figs.~\ref{fig:numerical_minus_infinity} --\ref{fig:numerical_plus_horizon_ratio}.
Results for MC orbits are shown in Figs.~\ref{fig:numerical_minus_infinity} and~\ref{fig:numerical_minus_horizon_ratio}. 
We take $M \oc=-10^{-3}$ (velocity profile in Fig.~\ref{fig:vel_prof}) and calculate fluxes in the dipole, quadrupole, and octupole modes. We will compare our results against relativistic Larmor formula RLF)~\cite{Jackson:1998nia},
\begin{equation}
\dot{E}^{\rm RLF}=\frac{2}{3} q^2 \frac{r_0^2 \Omega_0 ^4}{\left(1- r_0^2 \Omega_0 ^2\right)^2} \, .\label{eq:rel_larmor_circular}
\end{equation}
It is important to recall that this is a flat-space expression, and we will be feeding it with the curved-space result~\eqref{eq:omega_0} for $\Omega_0$, thus the comparison should be taken with a grain of salt. For large enough orbital radius $r_0$, MC orbits are relativistic, Sec.~\ref{sec:mag_bh}. Therefore, we expect higher order multipoles to be excited as a result of beaming~\cite{Jackson:1998nia}. This effect is already taken into account in the RLF~\eqref{eq:rel_larmor_circular}, which includes the contributions from all modes.

As we showed in Fig.~\ref{fig:vel_prof}, the charge velocity $r_0 \Omega_0$ is lowest for $r_0\sim 100M$. At this location, one could expect a non-relativistic approximation to describe the problem well. Thus, one would expect the dipolar fluxes to get closer to the total flux, an expectation that is apparent in our numerics. The analytical predictions~\eqref{eq:anres_ene_flux_inf_even} were derived on the assumption of slow motion, hence they agree best with our numerics also in a similar range of orbital radii. Nevertheless, the total flux predicted by the RLF~\eqref{eq:rel_larmor_circular} is never well approximated (to better than 10\% or so) by dipolar fluxes, and the discrepancy is largest for relativistic orbits, as expected.

Figure~\ref{fig:numerical_minus_infinity} also shows the total flux, in this case approximated by the sum of fluxes in the first three modes. As one can see, the sum of fluxes in the first three modes is in very good agreement with the RLF result~\eqref{eq:rel_larmor_circular} for a wide range of orbits, up to about $r_0\approx 350 M$, at which point the orbits become even more relativistic and so modes with $\ell >3$ would have to he included.

Finally, we also include a comparison with the energy flux one would get by neglecting the tail~\eqref{eq:sch_energy_change}. We see that for small orbital radius it cannot be neglected, as this results in a very large underestimation of the energy radiated; by contrast, for very large orbital radius, when orbits are dominated by magnetic effects, this expression recovers the RLF result, meaning the tail {\it can} be neglected in this case\footnote{To insist: even in this case, there is no ``orbital widening"}.

In Fig.~\ref{fig:numerical_minus_horizon_ratio} we look at energy fluxes into the horizon for particles in MC orbits, normalized by radiation flux at large distances. The first clear result is that radiation going into the horizon is in general highly suppressed, unless the charged particle is oribiting close to the BH. The dipolar flux dominates over all others, reaching about 10\% of the flux at infinity when the particle is at the ISCO. Our analytical predictions~\eqref{eq:anres_ene_flux_horizon_even} and~\eqref{eq:anres_ene_flux_inf_even} show very good agreement with with the numerical results, in this range of orbital radii.

We find that MC orbits can be subdivided in three regimes, defined with respect to the critical radius. These are, for $M\oc = -10^{-3}$ and $r_c = 100M$:

\begin{table}
    \centering
    \begin{tabular}{c|c|c|c}
       \hline
       $\ell$ & $m$ & $\dot{E}^\infty _{\ell m} / \dot{E}^\infty _{11}$ (\%) & $\dot{E}^H _{\ell m} / \dot{E}^\infty _{11}$ (\%) \\
       \hline
        1 & 1 & 100 & 4.50 \\
        2 & 2 & 38.4 & 0.19 \\
		2 & 1 & $<$ $10^{-1}$ & $<$ $10^{-2}$ \\
		3 & 3 & 13.1 & $<$ $10^{-2}$ \\
		3 & 2 & $<$ $10^{-1}$ & $<$ $10^{-3}$ \\
		4 & 4 & 4.31 & $<$ $10^{-3}$ \\
		5 & 5 & 1.39 & $<$ $10^{-5}$ \\
		6 & 6 & 0.44 & $<$ $10^{-6}$ \\
		7 & 7 & 0.14 & $<$ $10^{-8}$ \\
		8 & 8 & $<$ $10^{-1}$ & $<$ $10^{-9}$ \\
        \hline
    \end{tabular}
    \caption{Energy flux $\dot{E}^\infty _{\ell m}$ at infinity and at the horizon, $\dot{E}^H _{\ell m}$, for a particle orbiting the ISCO of a magnetized BH, such that $M \oc = - 10^{-3}$ in an MC configuration. We normalize these quantities by the dipolar flux $\dot{E}^\infty _{11}$. It is apparent that we can approximate the total energy flux with arbitrary precision using only finitely many modes.}
    \label{tab:numerical_minus_ISCO}
\end{table}
\noindent {\bf ``Kepler" region} $ r_0< 30 M$: orbits are highly relativistic and dominated by gravitational effects (Fig.~\ref{fig:vel_prof}). Our analytical predictions~\eqref{eq:anres_ene_flux_horizon_even} and~\eqref{eq:anres_ene_flux_inf_even} fail, as does the ``no tail" expression~\eqref{eq:sch_energy_change}, and not even the RLF is valid in this region. Adding the contributions from different modes calculated numerically, taking into account both the flux at infinity and on the horizon, we can still obtain an accurate value for the total energy flux. We confirm this for the first $(\ell \, , \, m)$ modes at $r_0 = r_\text{ISCO}$ in Table~\ref{tab:numerical_minus_ISCO}.

\noindent {\bf ``Slow orbit" region} $30M < r_0 < 180M$: these are the slowest orbits, with $v_0 < 0.2$, located around the critical radius $r_c =100M$ (see Fig.~\ref{fig:vel_prof}), where both gravitational and magnetic effects can be important. Here the dipole mode is dominant, but adding the energy flux in the first three $\ell$ modes visibly improves the agreement with the RLF~\eqref{eq:rel_larmor_circular}. Our analytical formulas~\eqref{eq:anres_ene_flux_horizon_even} and~\eqref{eq:anres_ene_flux_inf_even} give accurate predictions, while the ``no tail" expression~\eqref{eq:sch_energy_change} still underestimates the energy flux. For smaller $M\oc$ orbits in this region can be slower, and we can even have slow orbits dominated by magnetic effects. In that case, the tail term is negligible.
	
\noindent {\bf ``Magnetic" region} $r_0>180M$: these are highly relativistic orbits, dominated by magnetic effects. The dipolar flux alone cannot adequately describe the energy flux at infinity, meaning higher order multipoles are needed. We were able to get good agreement with the RLF~\eqref{eq:rel_larmor_circular} for a wide range of orbits with $r_0 \lesssim 350M$ using only the first three multipoles. However, beyond $r_0 \approx 350M$, orbits are too relativistic and higher order modes need to be included. The energy flux on the horizon is always negligible in this region. Our analytical predictions~\eqref{eq:anres_ene_flux_horizon_even} and~\eqref{eq:anres_ene_flux_inf_even} are not valid. Again, the tail term can be neglected, leading to very good agreement with the RLF.\footnote{This agreement is eventually spoiled by the redshift correction apparent in Eq.~\eqref{eq:omega_MC_2}.}

As was mentioned above, changing $M \oc$ leads to a quantitative change in the boundaries of these regions, but without any new qualitative feature.

\subsection{Plus configuration orbits: $\oc > 0$}
%
%
\begin{figure}
    \centering
    \includegraphics[width= .5 \textwidth]{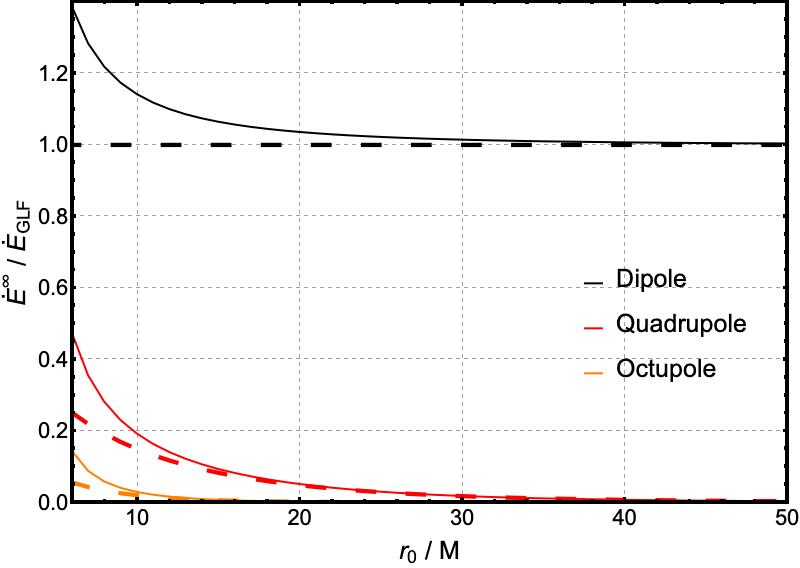}
   \caption{Energy flux of EM radiation at infinity, normalized to the GLF, for a charged particle in a PC circular orbit of radius $r_0$ around a BH with $M \oc=10^{-2} $. Solid curves correspond to numerical results, dashed lines represent the predictions obtained analytically~\eqref{eq:anres_ene_flux_inf_even}. Lowest orbital radius $r_0$ corresponds to the ISCO. We find good agreement between numerical and analytical results in the region $r_0>15 M$, where the dipole is dominant. }
    \label{fig:numerical_plus_infinity_larmor}
\end{figure}
We now turn to PC orbits. The velocity profile is shown in Fig.~\ref{fig:vel_prof} for $M \oc = 10^{-2}$, and the energy flux at infinity is shown in Fig.~\ref{fig:numerical_plus_infinity_larmor}.

In PC configurations, asymptotic orbits are slow and so the dipole dominates
the emission, as is clear in figure. This also means that our analytical results are valid, in particular the GLF expression~\eqref{eq:anres_ene_flux_inf}. Very close to the BH, however, the velocity is larger, and these approximations cease to be good descriptions. 

The ``no tail" expression of Eq.~\eqref{eq:sch_energy_change} predicts that all particles in orbits beyond the light ring, $r_0>3M$, gain energy ($\dot{E}<0$). As we already pointed out, this is wrong, and not borne out of our calculations.

%
\begin{figure}
    \centering
    \includegraphics[width= .5 \textwidth]{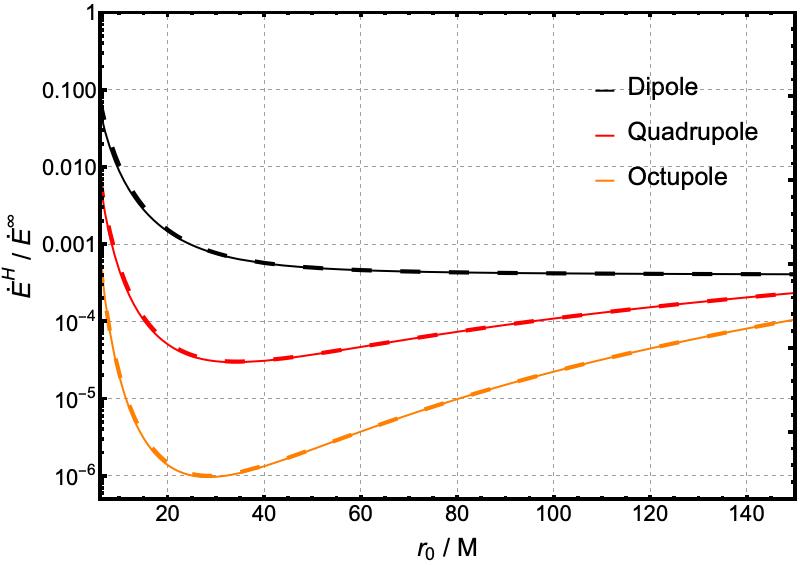}
    \caption{Ratio between the energy flux of EM radiation on the horizon and at infinity, for a charged particle in a PC circular orbit of radius $r_0$ around a Schwarzschild BH with $M \oc=10^{-2} $. Solid curves correspond to numerical results, dashed lines represent the predictions obtained analytically using~\eqref{eq:anres_ene_flux_horizon_even} and~\eqref{eq:anres_ene_flux_inf_even}. Lowest orbital radius $r_0$ corresponds to the ISCO. We find good agreement between numerical and analytical results. The dipole term still dominates, as does the energy flux on the horizon; however, for $r_0 /M > 40 $ we see the ratio going to a constant for the dipole, and starting to grow for the higher multipoles.}
    \label{fig:numerical_plus_horizon_ratio}
\end{figure}
%
The ratio of the energy flux at the horizon to that at infinity is shown in Fig.~\ref{fig:numerical_plus_horizon_ratio}. Our numerics show very good agreement with the slow-motion analytical results for $r_0\gtrsim 15M$. The dipole mode dominates emission.

As can be seen, the flux into the horizon is suppressed relative to radiation far away from the BH. Nevertheless, to our surprise, the ratio
$\dot{E}^H/\dot{E}^{\infty}$ does not decrease when the orbital radius increases: instead, it asymptotes to a constant. According to the analytical result of Eq.~\eqref{eq:anres_ratio_horizon_inf_PC}, the constant is $4 M^2 \oc^2 = 4 \times 10^{-4}$, which is consistent with the numerical results.
In fact, if we focus on a single multipole $\ell>1$, the ratio actually grows for larger orbital radius. This counter-intuitive finding (a radiating object gets farther from the BH, yet fluxes o
into the BH become more and more important) was already discussed in the context of our analytical, low-frequency approximations, Eq.~\eqref{eq:anres_ratio_horizon_inf_PC}. We dubbed it the ``horizon dominance effect," and as mentioned before, it is not a peculiarity of BH spacetimes: a flat calculation yields the same features (see Appendix \ref{app:flat_space}).

%
\section{Conclusions} \label{sec:conclusion}
%
In this work we studied a radiating charged particle in a circular orbit around a weakly magnetized Schwarzschild BH. The equation of motion governing the charge is the De Witt--Brehme equation~\eqref{eq:rad_motion_reduced}, which contains a complicated non-local tail term. Some literature~\cite{2018ApJ...861....2T,Tursunov:2018udx,Juraev:2024dju} claims that this term can be neglected in most cases for this system.
In a previous work, we had shown that the tail term must be included in a description of radiation in the Newtonian limit, unless the gravitational force is negligible in comparison to the Lorentz force~\cite{Santos:2023uka}.

The results presented here, both numerical and analytical, show that an identical conclusion holds true in the strong field regime, that is: \emph{in general, the tail term can only be neglected if the orbit is dominated by magnetic effects.} This dominance of magnetic effects can only happen for MC orbits, where the magnetic force is centripetal; for these orbits, the tail term can indeed be neglected if the orbital radius is much larger than the critical radius $r_c$, Eq.~\eqref{eq:critical_radius}. For PC orbits, where the magnetic force is centrifugal, the gravitational force is non-negligible for all orbital radii, and the tail term can never be neglected. By including the tail term in the study of PC orbits, we find that the ``orbital widening" effect can not occur. 

In the context of PC orbits, we found a surprising and counter-intuitive ``horizon dominance effect": the energy flux going into the horizon is a sizable fraction of that radiated to infinity. In fact, it approaches the nonzero constant $4 M^2 \oc^2$ as the orbital radius goes to infinity (and the particle velocity goes to zero), which can even be larger than unit: one can find a vast region of parameter space for which the fluxes into the horizon are larger than those at infinity. It is only natural to speculate that, once BH spin is included, the horizons fluxes can become negative, giving rise to outspiralling or to floating orbits~\cite{Cardoso:2011xi}.

\acknowledgments
%
We acknowledge support by VILLUM Foundation (grant no.\ VIL37766) and the DNRF Chair program (grant no.\ DNRF162) by the Danish National Research Foundation.
V.C.\ is a Villum Investigator and a DNRF Chair.  
J.S.S.\ and V.C.\ acknowledge financial support provided under the European Union’s H2020 ERC Advanced Grant “Black holes: gravitational engines of discovery” grant agreement no.\ Gravitas–101052587. 
Views and opinions expressed are however those of the author only and do not necessarily reflect those of the European Union or the European Research Council. Neither the European Union nor the granting authority can be held responsible for them.
This project has received funding from the European Union's Horizon 2020 research and innovation programme under the Marie Sklodowska-Curie grant agreement no.\ 101007855 and no.\ 101131233.
J.N.\ was partially supported by FCT/Portugal through CAMGSD, IST-ID, projects UIDB/04459/2020 and UIDP/04459/2020, and also by the H2020-MSCA-2022-SE project EinsteinWaves, GA no.\ 101131233.
%

%
%
%
%
%
\appendix
\section{Matched asymptotic expansions}  \label{app:matched_asymptotic_expansions}
%
%
%
To obtain the analytical results presented in Sec.~\ref{sec:analytical}, the main challenge is to solve the homogeneous Teukolsky radial equation (Eq.~\eqref{eq:sch_teukolsky_radial}) with physical boundary conditions. This remains true for other values of $s$, corresponding to scalar ($s=0$), gravitational ($s=-2$), or other types of radiation: the flux formulas do change, only the homogeneous equation to be solved. The boundary conditions are different for the two solutions $R^H _{\omega \ell m}(r)$ (purely ingoing waves at the horizon) and $R^\infty _{\omega \ell m} (r)$ (purely outgoing waves at infinity), which have the following asymptotic behavior~\cite{Teukolsky:1973ha}:
\begin{equation}
	R^H _{\omega \ell m} \sim  A_{\text{in}} \frac{e^{-i \omega r_\star}}{r} + A_{\text{out}} \frac{e^{i \omega r_\star}}{r^{2s+1}} \, , \quad (r \to \infty) \, ,
	\label{eq:app_asymptotic_hor_inf}
\end{equation}
\begin{equation}
	R^\infty _{\omega \ell m} \sim \frac{e^{i \omega r_\star}}{r^{2s+1}} \, , \quad (r \to \infty) \, ,
	\label{eq:app_asymptotic_inf_inf}
\end{equation}
\begin{equation}
	R^H _{\omega \ell m}\sim \frac{e^{-i \omega r_\star}}{\lr{r^2 f}^s} \, , \quad (r \to 2M) \, ,
	\label{eq:app_asymptotic_hor_hor}
\end{equation}
\begin{equation}
	R^\infty _{\omega \ell m} \sim B_\text{in} \frac{e^{-i \omega r_\star}}{\lr{r^2 f}^s} + B_\text{out} e^{i \omega r_\star} \, ,  \quad (r \to 2M) \, .
	\label{eq:app_asymptotic_inf_hor}
\end{equation}
If we take $s=-1$ in the equations above, we recover Eqs.~\eqref{eq:phi2_homo_ass_infty} and~\eqref{eq:phi2_homo_ass_2M}. The problem of finding analytical solutions to Teukolsky's equation is as old as the equation itself, starting with the work of Starobinsky and Churilov in 1974~\cite{Starobinskil:1974nkd}, but with developments in the following years~\cite{Page:1976df, 1986JMP....27.1238L, Mano:1996vt}. Here we will focus mainly on the procedure outlined in~\cite{Starobinskil:1974nkd, Page:1976df} to solve the homogeneous Teukolsky equation in the small frequency limit, $M |\omega| \ll 1$, where $\omega$ is the mode frequency and $M$ is the BH mass. We will calculate everything for generic $\ell$ and $s$, and then use these results in the main text to study the EM case ($s=-1$), as well and to briefly comment on the scalar ($s=0$) and gravitational ($s=-2$) cases in Appendix~\ref{app:scalar_grav}.
%
%
%
\subsection{The low-frequency limit of the homogeneous Teukolsky equation} 
%
%
%
Proceeding to the actual calculation, the first thing to do is solve the homogeneous Teukolsky equation, Eq.~\eqref{eq:sch_teukolsky_radial} with the right hand side set to zero, in the relevant limit. To do so, we start by defining
\begin{equation}
	x \equiv \frac{r}{2 M } - 1 \quad \text{and} \qquad \varpi \equiv 2 M \omega \, .
\end{equation}
In the $|\varpi| \ll 1$ regime, the radial homogeneous equation can be approximated by 
\begin{align}
    &x^2 (x+1)^2 \frac{d^2  R_{\omega \ell m}}{dx^2} \nonumber\\
    & + (s+1)x(x+1)(2 x +1) \frac{d R_{\omega \ell m}}{dx} \label{eq:A7}  \\
    &+\big[\varpi^2 x^4 + 2 i s \varpi x^3\nonumber \\
    & \qquad- (\ell-s)(\ell +s+1) x (x+1) - i s \varpi \big] R_{\omega \ell m} =0 \, .\nonumber
\end{align}
The idea now is to solve this equation for small and large $x$ and then match the two solutions. Starobinsky~\cite{Starobinskil:1974nkd} discovered that there is a similar equation, which differs from Eq.~\eqref{eq:A7} only by subdominant terms, and which is much easier to solve near the BH horizon \footnote{We compared the two solutions numerically and found their behavior to be indeed very similar.}. Therefore, the equation we will be using from now on is (dropping the subscripts in $R_{\omega \ell m}$):
\begin{align}
    &x^2 (x+1)^2 \frac{d^2  R}{dx^2} + (s+1)x(x+1)(2 x +1) \frac{d R}{dx}\label{eq:sch_teukolsky_radial_small_freq}  \\
    & + \big[\varpi^2 x^4 + 2 i s \varpi x^3 - (\ell-s)(\ell +s+1) x (x+1) \nonumber \\
    & \qquad \qquad \qquad \qquad \qquad  - i s \varpi (2 x +1) + \varpi^2 \big] R =0 \, .\nonumber
\end{align}
%
%
%
%
\subsubsection{Near region} 
%
%
%
To look at the region closer to the BH we consider the limit  $|\varpi| x \ll \ell (\ell +1)$. That means that we can neglect the first two terms inside the square brackets in~\eqref{eq:sch_teukolsky_radial_small_freq}, yielding 
\begin{align}
    &x^2 (x+1)^2 \frac{d^2  R}{dx^2} + (s+1)x(x+1)(2 x +1) \frac{d R}{dx} \label{eq:sch_teukolsky_radial_small_freq_near}\\
    &+\big[- (\ell-s)(\ell +s+1) x (x+1) \nonumber \\ 
    & \qquad \qquad  \qquad \quad \ \ - i s \varpi (2 x +1) + \varpi^2 \big] R =0 \, . \nonumber
\end{align}
Considering the ansatz 
\begin{equation}
	R (x) = x ^{-s - i \varpi} (1+x)^{-s+i\varpi} F(x)
\end{equation}
we obtain the solution
\begin{widetext}
\begin{align}
	R_{\text{near}} (x) 
	= &C_\text{in} \frac{x^{- i \varpi} (1+x)^{i \varpi}}{4^{s}\left[ x(1+x)\right]^s} \ _2 F_1 (- \ell - s, 1 + \ell - s; 1 - 2 i \varpi - s; -x) \label{eq:sch_teukolsky_radial_small_freq_near_solution} \\ 
	+& C_\text{out} \frac{x^{i \varpi} (x+1)^{i \varpi}}{(x+1)^{s}}  \, _2F_1(- \ell +2 i \varpi,1+ \ell +2 i \varpi;1+2 i \varpi+s;-x) \, ,\nonumber 
\end{align}
\end{widetext}
where $F$ is the hypergeometric function of the second kind. In this form, taking $C_\text{in} = 1$ and $C_\text{out} = 0$ yields the asymptotic behavior of $R^H$ at the horizon; for $R^\infty$, on the other hand, we can identify $C_\text{in}$ and $C_\text{out}$ with $B_\text{in}$ and $B_\text{out}$, respectively.
%
%
%
\subsubsection{Far region} 
%
%
%
To look at the region far away from the BH we must take $ x \gg \ell (\ell +1)$. This means that we can replace $(x+1)\to x $ and neglect the last two terms inside the square brackets in~\eqref{eq:sch_teukolsky_radial_small_freq} altogether, obtaining 
\begin{align}
    &x^4 \frac{d^2  R}{dx^2} + 2(s+1)x^3 \frac{d R}{dx} \label{eq:sch_teukolsky_radial_small_freq_far} \\ 
    &+ \big[\varpi^2 x^4 + 2 i s \varpi x^3  - (\ell-s)(\ell +s+1) x ^2  \big] R =0 \, . \nonumber
\end{align}
The solution to this equation is a combination of confluent hypergeometric functions. After some manipulation, we can write it as  
\begin{widetext}
	\begin{align}
	R_\text{far}(x) 
	=& C_3\ x^{\ell -s} e^{- i \varpi x} \ U(\ell-s+1,2 \ell +2,2 i \varpi x) + C_4 \ x^{\ell -s} e^{- i \varpi x} \ M(\ell-s+1;2 \ell +2;2 i \varpi x) \, ,\label{eq:sch_teukolsky_radial_small_freq_far_solution}
\end{align}
\end{widetext}
where $M$ and $U$ are confluent hypergeometric functions of the first and second kinds, respectively. Next, we focus on matching the solutions of~\eqref{eq:sch_teukolsky_radial_small_freq_near_solution} and~\eqref{eq:sch_teukolsky_radial_small_freq_far_solution} in the region where $\ell (\ell +1) \ll x \ll \ell (\ell +1) / |\varpi|$. We will do this separately for the two relevant boundary conditions, purely ingoing waves at the horizon and purely outgoing waves at infinity. 

Before doing so, however, we recall that our goal is to compute the result of Eq.~\eqref{eq:sch_teukolsky_Zlm_inf}. Thus we must find $A_\text{in}$ and $R^{H, \, \infty}(x_0)$, where $x_0 = r_0 / 2M -1 $. Finding $A_\text{in}$ is straightforward and is done in the next subsection, but to find $R^{H, \, \infty}(x_0)$ we must determine which branch of the analytical solution is valid at $x_0$. Since we are interested in studying circular orbits and we are working in the $|\varpi| \ll 1$ limit, we choose to use the near region branch, which is valid for studying slow orbits, as $\omega = m \Omega_0$, and thus
\begin{equation}
	 |v_0|  \ll 1 < \frac{\ell  (\ell +1)}{|m|} \implies |\varpi| x_0  \ll \ell (\ell +1) \, ,
\end{equation}
where $\Omega_0$ and $v_0 = r_0 \Omega_0$ are the angular and linear velocities measured by a stationary observer at infinity, respectively.
%
%
%
\subsection{Matching: purely ingoing waves at the horizon} 
%
%
%
The solution with purely ingoing boundary conditions at the horizon is $R^H$, and we want it to exhibit the asymptotic behavior given in Eqs.~\eqref{eq:app_asymptotic_hor_inf} and~\eqref{eq:app_asymptotic_hor_hor}. Solving for $R^H$ will allow us to obtain two of the quantities we are looking for: $A_\text{in}$ and $R^H (x_0) = R^H _\text{near} (x_0) $. The first step is to impose the behavior at the horizon, which is done by setting $C_\text{in}=1$ and $C_\text{out}=0$ in~\eqref{eq:sch_teukolsky_radial_small_freq_near_solution}, so that 
\begin{align}
	R^H _\text{near} = &\frac{x^{- i \varpi} (1+x)^{i \varpi}}{4^{s}\left[ x(1+x)\right]^s} \label{eq:sch_teukolsky_radial_small_freq_near_solution_hor} \\
	& \quad \times \ _2 F_1 (- \ell - s, 1 + \ell - s; 1 - 2 i \varpi - s; -x) \, . \nonumber
\end{align}
Proceeding to larger values of $r$, the matching of the two solutions is done in the region where $\ell (\ell +1) \ll x \ll \ell (\ell +1)/ |\varpi|$. With that in mind, we start by looking at the $x \gg \ell (\ell +1)$ limit of~\eqref{eq:sch_teukolsky_radial_small_freq_near_solution_hor}. Ater some maniuplation we find Eq.~\eqref{eq:sch_teukolsky_radial_small_freq_near_solution_hor_far}. 

Now, turning to the behavior of~\eqref{eq:sch_teukolsky_radial_small_freq_far_solution} in the matching region where $x\ll \ell (\ell +1)/|\varpi|$, we must simply take the $ \varpi x \to 0 $ limit of this expression, which yields Eq.~\eqref{eq:sch_teukolsky_radial_small_freq_far_solution_hor_near}. It is easy to see that both expressions have the same dependency on $x$, and so the matching is done by looking at the coefficients. Thus, to match the solutions we must have $C_3$ and $C_4$ given by Eqs.~\eqref{eq:app_match_hor_c3} and~\eqref{eq:app_match_hor_c4}, respectively. 

Besides determining the value of $R^H$ at the particle's orbit, we also want to find the coefficient for the ingoing mode at infinity, $A_\text{in}$ (see Eq.~\eqref{eq:app_asymptotic_hor_inf}). This is done by looking at the $x\to \infty$ limit of $R^H_\text{far}$, yielding the result in Eq.~\eqref{eq:app_match_hor_ain}.
\begin{widetext}
\begin{align}
	R^H _{\text{near}} \sim 
	&  \,\, i \,  \varpi (-1)^{s} 2^{1-2 s} \Gamma (1-s-2 i \varpi) \frac{ \ell ! \, (\ell +s)!}{(2 \ell +1)!} x^{-1- \ell-s} \label{eq:sch_teukolsky_radial_small_freq_near_solution_hor_far}  + \frac{4^{-s} \Gamma (2 \ell +1) \Gamma (1-s-2 i \varpi)}{\Gamma (\ell-2 i \varpi+1) \Gamma (\ell-s+1)} x^{\ell -s} \quad \lr{x\gg \ell (\ell +1)} \, , \\
	R^H _\text{far}
	\approx & \left(\frac{\Gamma (-1-2l)}{\Gamma (- \ell -s)}C_3  + C_4 \right) x^{\ell -s} - \left(\frac{i \, \ell \, e^{-i \pi  l}}{ 4^{\ell } 
   \varpi^{2 \ell +1} }\frac{ \Gamma (2 l)}{\Gamma (\ell-s+1)} C_3 \right) x^{- \ell-s-1}  \quad (|\varpi| x\ll \ell (\ell +1)) \, ,\label{eq:sch_teukolsky_radial_small_freq_far_solution_hor_near} \\ \nonumber \\ \nonumber \\
	C_3 =& \frac{e^{i \pi  l} (-1)^{s+1} \varpi^{2 \ell +2} 2^{2 l-2 s+1}  \Gamma (\ell-s+1) \Gamma (-2 i \varpi-s+1)\,  \ell ! \, (\ell +s)!}{\ell  \,  \Gamma (2 l) \, (2 \ell +1)!}	\, ,\label{eq:app_match_hor_c3} \\ \nonumber \\
	C_4 = & \frac{(-1)^{- \ell} 2^{-2 s} \Gamma (-2 i \varpi-s+1)}{\, \ell ((2 \ell +1)!)^2 \Gamma (2 l) \Gamma (\ell-2 i \varpi+1) \Gamma (\ell-s+1)} \label{eq:app_match_hor_c4} \\
		& \times \Big((-1)^l \Gamma (2 l) \Gamma (2 \ell +1) \, \ell \, ((2 \ell +1)!)^2 + 2^{2 \ell +1} e^{i \pi  l} (-1)^{2 s+1} \varpi^{2
   \ell +2}  \Gamma (\ell-2 i \varpi+1) \Gamma (\ell-s+1)^2 \,  \ell ! \, ((\ell +s)!)^2 \Big) .
   \nonumber \\ \nonumber \\ \nonumber \\ 
	A_\text{in} = & -\frac{2^{-l-s} e^{i \pi  (l-s)} \Gamma (2 l+1) \Gamma (2 l+2) (i \varpi)^{-l+s-1} \Gamma (-2 i \varpi-s+1)}{\Gamma (l-2 i \varpi+1) \Gamma (l-s+1) \Gamma
   (l+s+1)}
	\label{eq:app_match_hor_ain}
\end{align}
\end{widetext}
Thus, to compute the energy flux to infinity, we already have everything we need in $A_\text{in}$ and $R^H _\text{near}(x)$; however, to obtain the energy flux into the horizon, we still need to find $R^\infty _\text{near}(x)$. 
%
%
%
\subsection{Matching: purely outgoing waves at infinity} 
%
%
%
The procedure to perform the matching and find $R^\infty (x)$ is similar to that of the previous subsection, except that now the boundary conditions are given by Eqs.~\eqref{eq:app_asymptotic_inf_inf} and~\eqref{eq:app_asymptotic_inf_hor}. That sets $C_3$ and $C_4$, and once we have the far region solution we can match it to the near region one to obtain $R^\infty _\text{near}(x)$, which is precisely what we need for Eq.~\eqref{eq:sch_teukolsky_Zlm_inf}, as $R^\infty(r_0) = R^\infty _\text{near} (x_0)$. The matching itself is done in an identical manner to the previous subsection.

With that in mind, for conciseness, we only show the relevant coefficients $C_\text{in}$ and $C_\text{out}$, which go into the near solution:
\begin{widetext}
\begin{align}
	C_\text{in} =& -\frac{2^{-l+s-2} \varpi^{-2 (l+1)} (i \varpi)^{l+s+1} \Gamma (l-2 i \varpi+1) \Gamma (l-s+1)}{\Gamma (2
   l+1) \Gamma (2 l+2) | l| ! | 2 l+1| ! \Gamma (l+s+1) \Gamma (-2 i \varpi-s+1)} \label{eq:app_match_inf_cin} \\ 
   &\times \frac{1}{(-1)^{2 s} \Gamma (l-2 i \varpi+1) \Gamma (l-s+1) | l+s| !+\Gamma (l+2
   i \varpi+1) \Gamma (l+s+1) | l-s| !} \nonumber \\
   & \times \Big(\Gamma (2 l+1)^2 \Gamma (2 l+2) (| 2 l+1| !)^2+4^{l+1}
   \varpi^{2 l+2} | l| ! \Gamma (l+2 i \varpi+1) \Gamma (l-s+1) \Gamma (l+s+1) | l \nonumber \\
   & \hspace{6.7cm} -s| ! (\Gamma (2 l+2) | l+s| !-| 2 l+1| ! \Gamma (l+s+1))\Big) \, , \nonumber \\
	C_\text{out} =& \frac{\varpi^{-2 (l+1)} 2^{-l-s-2} (i \varpi)^{l+s+1} \Gamma (l+2 i \varpi+1)}{\Gamma (2 l+1) \Gamma
   (2 l+2) | l| ! | 2 l+1| ! \Gamma (2 i \varpi+s+1)} \label{eq:app_match_inf_cout} \\
   &\times \frac{1}{ e^{2 i \pi  s} \Gamma (l-2 i \varpi+1) \Gamma (l-s+1) | l+s| !+\Gamma (l+2 i \varpi+1) \Gamma
   (l+s+1) | l-s| !} \nonumber \\ 
   &\times \Big(\Gamma (2 l+1)^2 \Gamma (2 l+2) (| 2 l+1| !)^2-| 2 l+1| ! \Gamma (l+s+1)) \nonumber \\
   & \hspace{2 cm} -4^{l+1} e^{2 i \pi  s}
   \varpi^{2 l+2} | l| ! \Gamma (l-2 i \varpi+1) \Gamma (l-s+1)^2 | l+s| ! (\Gamma (2 l+2) | l+s| ! \Big)\, . \nonumber
\end{align}
\end{widetext}

We now know $R^\infty _\text{near} (x)$, and so we have completed our task. To obtain an analytic expression for the solution of the homogeneous Teukolsky equation with the boundary conditions discussed in the beginning of this Appendix, we just have to replace the desired values of $s$, $\ell$ and $m$, provided, of course, that the orbit in question satisfies $M |\omega| \ll 1$ as well as $|v_0| \ll 1$.
%
%
%
\section{Absorbing sphere in flat space}  \label{app:flat_space}
%
%
%
In this Appendix we study the flat space analogue of a particle orbiting a BH and radiating scalar waves. By this we mean a particle with scalar charge $\gamma$ which is in a circular orbit, in flat space, around an absorbing sphere of radius $r_1$. The equation we need to solve is just the Klein--Gordon equation with a source ,
\begin{equation}
    \square \Phi = \gamma T^M \, , 
    \label{eq:KG}
\end{equation}
where $\square$ is the D'Alembertian operator in flat space, $\Phi$ is the scalar field, and $T^M$ is the trace of the particle's energy-momentum tensor. In the case of the particle in a circular orbit, the energy-momentum tensor is
\begin{align}
    T^M_{\mu \nu} = &\int d \tau \, u_\mu (\tau) u_\nu (\tau) \delta^4 (x - z(\tau)) \label{eq:T_matter}\\ 
    =& \frac{1}{u^t}  u_\mu (t) u_\nu (t) \frac{\delta(r-r_0)}{r^2} \frac{\delta(\theta - \pi/2)}{\sin \theta} \delta(\phi - \Omega_0 t) \nonumber \, ,
\end{align}
where the notation is the same as in Eq.~\eqref{eq:Jmu}. 

Just as before, we want to study the energy flux of scalar waves at infinity and on the surface of the absorbing sphere. The method is identical to what was done in the main text for electromagnetic waves in the Schwarzschild geometry. We start by decomposing the field and sources in Fourier harmonic components, %
\begin{equation}
    \Phi = \int d\omega \sum_{\ell m} R_{\omega \ell m}(r) Y_{\ell m}(\theta , \phi) e^{- i \omega t} \, ,
    \label{eq:FH_Phi}
\end{equation}
\begin{equation}
    T^\text{M} = \int d\omega \sum_{\ell m} T_{\omega \ell m}(r) Y_{\ell m}(\theta , \phi) e^{- i \omega t} \, ,
    \label{eq:FH_T}
\end{equation}
thus obtaining a decoupled radial equation
\begin{align}
    \left( \frac{d^2 }{dr^2} + \frac{2}{r} \frac{d}{dr} + \omega^2 - \frac{\ell(\ell+1)}{r^2} \right) R_{\omega \ell m} = \gamma \, T_{\omega \ell m} \, .
    \label{eq:scalar_radial}
\end{align}
The solution of this equation can again be written in the form of Eq.~\eqref{eq:gen_sol_r}, with $s=0$. The homogeneous equation (put $T_{\omega \ell m}=0$ above) is just the spherical Bessel equation, so the solutions can be expressed as a combination of spherical Hankel functions, 
\begin{equation}
    h_\ell ^{(1,2)}(x) = \mp i (-x)^\ell \left(\frac{1}{x} \frac{d }{dx}\right)^\ell \left( \frac{ e^{\pm i x}}{x}\right) \, ,
\end{equation}
where $x= \omega r$. We now must define $R^H$ and $R^\infty$, the solutions of the homogeneous equation corresponding to ingoing waves at $r_1$ and outgoing waves at infinity, respectively. By looking at the behavior of $h_\ell ^{(1,2)}(\omega r)$ for large $r$, we find 
\begin{equation}
   	h_\ell ^{(1,2)}(\omega r) \sim \frac{e^{\pm i \omega r}}{r} \, , \quad (\omega r \gg 1) \, .
\end{equation}
Thus, for the solution corresponding to purely outgoing waves at infinity we have simply 
\begin{equation}
    R^\infty_{\omega \ell m} (r)= 	h_\ell ^{(1)}(\omega r) \, .
\end{equation}
As for the absorbing condition at $r_1$, the situation is a bit more complicated. In the BH case, the horizon is a null hypersurface, where it is possible to define ingoing waves in an unambiguous manner. A spherical surface in flat space does not define a null hypersurface, and it is not easy to tell wether a given function corresponds to an ingoing or outgoing wave for any finite radius $r_1$. The only reasonable alternative is to look directly at the energy flux, and force it to be ingoing on the absorbing surface. The energy-momentum tensor of the scalar field is
\begin{equation}
    T^\text{SF}_{\mu \nu} = \frac{1}{2} \left( \partial_\mu \Phi \partial_\nu \overbar{\Phi} +  \partial_\mu \overbar{\Phi} \partial \nu \Phi  - \eta_{\mu \nu} \eta^{\alpha \beta} \partial_\alpha \Phi \partial_\beta \overbar{\Phi} \right) \, ,
    \label{eq:Tmunu}
\end{equation}
where $\eta_{\mu \nu}$ is the Minkowski metric. The ingoing energy flux on a spherical surface of radius $R$ is 
\begin{equation}
    \frac{d E}{d t } = - \int d\Omega \, R^2\,  T^\text{SF} _{r t } \, ,
    \label{eq:dEdt}
\end{equation}
From Eq.~\eqref{eq:Tmunu} we find that we can impose the energy flux to be negative if we require
\begin{equation}
    \partial_r \Phi (r_1)= k \, \partial_t \Phi (r_1) \, , \quad k \in \mathbb{R}^+ .
\end{equation}
For simplicity, and also by analogy with the absorbing boundary conditions in mumerical relativity \cite{Bayliss:1980, Sarbach:2007hd}, we choose $k=1$, which yields
\begin{equation}
   R^H_{\omega \ell m}(r) = h^{(1)}_\ell (\omega r) + \alpha_{\omega \ell m} h^{(2)} (\omega r)
    \label{eq:RH3}
\end{equation}
with
\begin{equation}
    \alpha_{\omega \ell m} = \frac{\omega r_1  h_{\ell+1}^{(1)}(\omega r_1 )-(\ell+i \omega r_1) h_\ell^{(1)}(\omega r_1 )}{-\omega r_1 
   h_{\ell+1}^{(2)}(\omega r_1 )+(\ell+i \omega r_1 )
   h_\ell^{(2)}(\omega r_1 )} \, .
\end{equation} 
Replacing these functions $R^{H,\infty} _{\omega \ell m}$, as well as the Fourier harmonic decomposition $T_{\omega \ell m}$ of the source term in Eq.~\eqref{eq:T_matter}, in the solution given in Eq.~\eqref{eq:gen_sol_r}, and then plugging that into \eqref{eq:FH_Phi}, yields
\begin{align}
    \Phi(x) = \sum_{\ell m} \Big[& R^\infty _{m \Omega_0 \ell m} (r) Z_{\ell m}^\infty \Theta(r-r_0) \label{eq:phi2_appendix} \\
    +& R^H_{m \Omega_0 \ell m} ( r) Z_{\ell m}^H \Theta(r_0-r) \Big] Y_{l m} e^{- i m \Omega_0 t} \, , \nonumber
\end{align}
where $\Theta(x)$ denotes the Heaviside step function and 
\begin{align}
    Z_{\ell m}^{\infty,H} =& i m \gamma \Omega_0 \sqrt{1 - (r_0 \Omega_0 )^2} \, R^{H,\infty} _{m \Omega_0 \ell m } (r_0) \ \overbar{Y}_{\ell m } (\pi/2, 0) \, .
\end{align}
Finally, this must be replaced in Eq.~\eqref{eq:dEdt} to find the energy flux at $R=\infty$, 
\begin{equation}
    \dot{E}^\infty  =\sum_{l m}\left| Z_{\ell m}^\infty \right|^2 \, , 
\end{equation}
and at $R=r_1$,
\begin{equation}
    \dot{E}^{r_1}  =\sum_{l m}\mathcal{A}_{\ell m} \left|  Z_{\ell m}^H \right|^2 \, , 
\end{equation}
where
\begin{widetext}
\begin{equation}
    \mathcal{A}_{\ell m} = -\frac{4}{(-m \Omega_0 r_1  h_{\ell+1}^{(1)}(m r_1
   \Omega_0)+(\ell-i m \Omega_0 r_1 ) h_\ell^{(1)}(m
   r_1 \Omega_0)) (-m \Omega_0 r_1 
   h_{\ell+1}^{(2)}(m \Omega_0 r_1 )+(\ell+i m r_1
   \Omega_0) h_\ell^{(2)}(m \Omega_0 r_1 ))} \, .
\end{equation}
\end{widetext}
In particular, we can look at the result for small velocity, $r_0 \Omega_0 \ll 1$. We find that in this limit, as expected, the dominant mode is the dipole mode, and we have, for the fluxes at infinity and $r_1$
\begin{align}
    \dot{E}^\infty _{1\,1}  \approx & \frac{ \gamma^2 (2 r_0 ^2 + r_1^2)^3 \Omega_0^4}{48 \pi r_0 ^4} \, , \label{eq:flat_ene_flux_inf} \\
    \dot{E}^{r_1}   _{1\,1} \approx &- \frac{3 \gamma^2 r_1 ^4 \Omega_0 ^2}{16 \pi r_0^4 } \, , 
\end{align}
respectively. Looking at the ratio of these two quantities and taking also the limit of large orbital radius yields
\begin{equation}
    \frac{ \dot{E}_{11}^{r_1}}{ \dot{E}_{11}^\infty} = -\frac{9 r_1^4}{4 r_0^6 \Omega_0 ^2} \, .
\end{equation}
This result recovers the scaling we had obtained for the ratio of energy flux on the horizon and at infinity for a particle orbiting a Schwarzschild BH in Eq.~\eqref{eq:anres_ratio_horizon_inf_B0}. The numerical factor is different, but this may be attributable to the fact that the absorbing boundary condition at $r_1$ does not imply perfect absorption. 

In general, this simple model for an absorbing sphere in flat space, which aims to replicate the boundary conditions on the horizon of a BH, recovers the ``horizon dominance effect" we reported in Sec.~\ref{sec:analytical}: for a given value of orbital radius, decreasing the orbital velocity causes the energy absorbed by the BH (or absorbing sphere in flat space) to become arbitrarily larger than the energy escaping to infinity (although, of course, the total energy radiated does go to zero).
%
%
%
%
\section{The scalar and gravitational cases} \label{app:scalar_grav}
%
%
%
As stated above, the analytical methods we developed can be used for any value of spin parameter. To do so, we just have to find the corresponding equivalent of Eqs.~\eqref{eq:sch_teukolsky_Zlm_inf},~\eqref{eq:sch_teukolsky_phi2_ene_flux_inf}, and~\eqref{eq:sch_teukolsky_phi2_ene_flux_hor}, which is done in Refs.~\cite{Cardoso:2019nis, Poisson:1993vp, Poisson:1994yf, Yunes:2011aa, Teukolsky:1974yv}. This allows us to do an analysis as complete as was done in the EM case, but here we only show the dominant modes for conciseness.
%
%
\subsection{Scalar radiation} \label{sec:scalar}
%
%
%
For the case of a particle coupling to a massless complex scalar field, we consider the action~\cite{Cardoso:2019nis}
\begin{equation}
    S = \int d^4 x  \sqrt{- g } \lr{\frac{R}{16 \pi} - g^{\mu \nu} \del _\mu \Phi \del_\nu \overbar{\Phi} - 2 \gamma \Phi T^M} \, ,
 \label{eq:scalar_action}
\end{equation}
where $\Phi$ is the scalar field, $g_{\mu \nu}$ is the metric, $g$ is the determinant of the metric, $R$ is the Ricci scalar, $\gamma$ is a coupling constant and $T^M$ is the trace of the stress energy tensor of the particle, which is taken to have mass $m_0$. If we treat the scalar field as a test field then its equation of motion is just the massless Klein-Gordon equation in the background geometry; if moreover we fix the later to be given by the Schwarzschild metric, we simply obtain Eq.~\eqref{eq:teukolsky} with $s=0$. Thus, we can solve this problem in an almost identical manner to what we did for the electromagnetic case, keeping in mind that the expressions for the energy flux in terms of the fields are different~\cite{Teukolsky:1974yv}. We find that the energy fluxes, in the dipole mode, at infinity and on the horizon are given by
\begin{align}
	&_{0} \dot{E}_ {11}  ^\infty +\, _{0} \dot{E}_ {1 -1}  ^\infty  = \frac{\gamma ^2 m_0^2}{12 \pi} \frac{ (r_0-2 M) (r_0-M)^2}{ r_0} \Omega_0^4 \, ,  \label{eq:scalar_ene_flux_inf} \\
	& _{0} \dot{E}_ {11}  ^H  = \frac{3 \gamma ^2m_0^2 (r_0-2 M) \Omega_0^2}{8 \pi M^2 r_0}  \label{eq:scalar_ene_flux_hor} \\ 
    & \qquad \qquad\times\left((r_0-M) \log \left(\frac{r_0}{r_0-2 M}\right)-2 M\right)^2 \, , \nonumber
\end{align}
respectively (these expressions are valid only for slow orbits $r_0 \Omega_0 \ll 1$). If we compare the two for large orbital radius, as we did in Eq.~\eqref{eq:anres_ene_flux_horizon}, we find 
\begin{equation}
\frac{_{0} \dot{E}  _{11}^H }{ _{0}\dot{E}_{11} ^\infty } \sim \frac{4 M^4}{r_0^6 \Omega_0 ^2 } \qquad (r_0 \to \infty) \, ,
	\label{eq:scalar_ene_flux_hor_inf}
\end{equation}
which exactly recovers the behavior we had found for EM radiation. 
%
%
\subsection{Gravitational radiation} \label{sec:grav}
%
%
%
As for the more studied case of gravitational waves, we can once again follow Ref.~\cite{Cardoso:2019nis}. Now the equation of motion is Eq.~\eqref{eq:teukolsky} with $s=-2$, but the method is almost the same, although the calculations and final expressions are more cumbersome. For a particle of mass $m_0$, we find the energy fluxes, in the quadrupole mode, at infinity and on the horizon, to be
\begin{widetext}
\begin{align}
	_{-2}\dot{E}_ {22}  ^\infty + _{-2}\dot{E}_ {2 -2}  ^\infty  = & \frac{32 m_0^2 (r_0-2M)^2 r_0 (r_0 (9 r_0-20 M)+36 M^2) \Omega_0^6}{45 (r_0 -3 M)} \label{eq:grav_ene_flux_inf} \, , 
\end{align}
\begin{align}
		_{-2}\dot{E}_ {22}  ^H  = 
		\frac{5 m_0^2 \Omega_0^2}{144 (r_0-3 M) r_0^4} \Bigg[& \frac{4}{M^2} \left(81 r_0^7-342 r_0^6 M +657 r_0^5 M^2 -588 r_0^4 M^3 -180 r_0^3 M^4 +400 r_0^2 M^5 +164 r_0 M^6 +64 M^7 \right) \nonumber \\
		&-\frac{12 r_0^3 }{M^3}    \left(27 r_0^5-141 r_0^4 M +324 r_0^3 M^2 -386 r_0^2 M^3 +104 r_0 M^4 +136 M^5\right) \log \left(\frac{r_0}{r_0-2M}\right) \nonumber \\
		&+ \frac{9 r_0^5}{M^4} (r_0-2M)^2 \left(9 r_0^2-20 r_0 M+36M^2\right)  \log^2\left(\frac{r_0}{r_0-2M}\right)\Bigg] \, , \label{eq:grav_ene_flux_hor}
\end{align}
\end{widetext}
respectively. If we compare the two for large orbital radius ($r_0 \to \infty$), we find 
\begin{equation}
\frac{_{-2}\dot{E} _{22}^H }{ _{-2} \dot{E}_{22}  ^\infty }  \sim \frac{M^6}{r_0 ^{10} \Omega_0^4}
	 \qquad (r_0 \to \infty) \, ,
	\label{eq:grav_ene_flux_hor_inf}
\end{equation}
which recovers standard literature results for Keplerian orbits \cite{Poisson:1994yf}. Moreover, even though the result is different from the scalar and electromagnetic cases, because the dominant mode is now the quadrupole rather than the dipole, we still find that the ``horizon dominance effect" can occur.  
%
%
%
%

\begin{thebibliography}{43}%
\makeatletter
\providecommand \@ifxundefined [1]{%
 \@ifx{#1\undefined}
}%
\providecommand \@ifnum [1]{%
 \ifnum #1\expandafter \@firstoftwo
 \else \expandafter \@secondoftwo
 \fi
}%
\providecommand \@ifx [1]{%
 \ifx #1\expandafter \@firstoftwo
 \else \expandafter \@secondoftwo
 \fi
}%
\providecommand \natexlab [1]{#1}%
\providecommand \enquote  [1]{``#1''}%
\providecommand \bibnamefont  [1]{#1}%
\providecommand \bibfnamefont [1]{#1}%
\providecommand \citenamefont [1]{#1}%
\providecommand \href@noop [0]{\@secondoftwo}%
\providecommand \href [0]{\begingroup \@sanitize@url \@href}%
\providecommand \@href[1]{\@@startlink{#1}\@@href}%
\providecommand \@@href[1]{\endgroup#1\@@endlink}%
\providecommand \@sanitize@url [0]{\catcode `\\12\catcode `\$12\catcode `\&12\catcode `\#12\catcode `\^12\catcode `\_12\catcode `\%12\relax}%
\providecommand \@@startlink[1]{}%
\providecommand \@@endlink[0]{}%
\providecommand \url  [0]{\begingroup\@sanitize@url \@url }%
\providecommand \@url [1]{\endgroup\@href {#1}{\urlprefix }}%
\providecommand \urlprefix  [0]{URL }%
\providecommand \Eprint [0]{\href }%
\providecommand \doibase [0]{http://dx.doi.org/}%
\providecommand \selectlanguage [0]{\@gobble}%
\providecommand \bibinfo  [0]{\@secondoftwo}%
\providecommand \bibfield  [0]{\@secondoftwo}%
\providecommand \translation [1]{[#1]}%
\providecommand \BibitemOpen [0]{}%
\providecommand \bibitemStop [0]{}%
\providecommand \bibitemNoStop [0]{.\EOS\space}%
\providecommand \EOS [0]{\spacefactor3000\relax}%
\providecommand \BibitemShut  [1]{\csname bibitem#1\endcsname}%
\let\auto@bib@innerbib\@empty
\bibitem [{\citenamefont {Dirac}(1938)}]{Dirac:1938nz}%
  \BibitemOpen
  \bibfield  {author} {\bibinfo {author} {\bibfnamefont {P.~A.~M.}\ \bibnamefont {Dirac}},\ }\href {\doibase 10.1098/rspa.1938.0124} {\bibfield  {journal} {\bibinfo  {journal} {Proc. Roy. Soc. Lond. A}\ }\textbf {\bibinfo {volume} {167}},\ \bibinfo {pages} {148} (\bibinfo {year} {1938})}\BibitemShut {NoStop}%
\bibitem [{\citenamefont {DeWitt}\ and\ \citenamefont {Brehme}(1960)}]{DEWITT1960220}%
  \BibitemOpen
  \bibfield  {author} {\bibinfo {author} {\bibfnamefont {B.~S.}\ \bibnamefont {DeWitt}}\ and\ \bibinfo {author} {\bibfnamefont {R.~W.}\ \bibnamefont {Brehme}},\ }\href {\doibase https://doi.org/10.1016/0003-4916(60)90030-0} {\bibfield  {journal} {\bibinfo  {journal} {Annals of Physics}\ }\textbf {\bibinfo {volume} {9}},\ \bibinfo {pages} {220} (\bibinfo {year} {1960})}\BibitemShut {NoStop}%
\bibitem [{\citenamefont {Hobbs}(1968)}]{HOBBS1968141}%
  \BibitemOpen
  \bibfield  {author} {\bibinfo {author} {\bibfnamefont {J.}~\bibnamefont {Hobbs}},\ }\href {\doibase https://doi.org/10.1016/0003-4916(68)90231-5} {\bibfield  {journal} {\bibinfo  {journal} {Annals of Physics}\ }\textbf {\bibinfo {volume} {47}},\ \bibinfo {pages} {141} (\bibinfo {year} {1968})}\BibitemShut {NoStop}%
\bibitem [{\citenamefont {Smith}\ and\ \citenamefont {Will}(1980)}]{Smith:1980tv}%
  \BibitemOpen
  \bibfield  {author} {\bibinfo {author} {\bibfnamefont {A.~G.}\ \bibnamefont {Smith}}\ and\ \bibinfo {author} {\bibfnamefont {C.~M.}\ \bibnamefont {Will}},\ }\href {\doibase 10.1103/PhysRevD.22.1276} {\bibfield  {journal} {\bibinfo  {journal} {Phys. Rev. D}\ }\textbf {\bibinfo {volume} {22}},\ \bibinfo {pages} {1276} (\bibinfo {year} {1980})}\BibitemShut {NoStop}%
\bibitem [{\citenamefont {Mino}\ \emph {et~al.}(1997)\citenamefont {Mino}, \citenamefont {Sasaki},\ and\ \citenamefont {Tanaka}}]{Mino:1996nk}%
  \BibitemOpen
  \bibfield  {author} {\bibinfo {author} {\bibfnamefont {Y.}~\bibnamefont {Mino}}, \bibinfo {author} {\bibfnamefont {M.}~\bibnamefont {Sasaki}}, \ and\ \bibinfo {author} {\bibfnamefont {T.}~\bibnamefont {Tanaka}},\ }\href {\doibase 10.1103/PhysRevD.55.3457} {\bibfield  {journal} {\bibinfo  {journal} {Phys. Rev. D}\ }\textbf {\bibinfo {volume} {55}},\ \bibinfo {pages} {3457} (\bibinfo {year} {1997})},\ \Eprint {http://arxiv.org/abs/gr-qc/9606018} {arXiv:gr-qc/9606018} \BibitemShut {NoStop}%
\bibitem [{\citenamefont {Quinn}\ and\ \citenamefont {Wald}(1997)}]{Quinn:1996am}%
  \BibitemOpen
  \bibfield  {author} {\bibinfo {author} {\bibfnamefont {T.~C.}\ \bibnamefont {Quinn}}\ and\ \bibinfo {author} {\bibfnamefont {R.~M.}\ \bibnamefont {Wald}},\ }\href {\doibase 10.1103/PhysRevD.56.3381} {\bibfield  {journal} {\bibinfo  {journal} {Phys. Rev. D}\ }\textbf {\bibinfo {volume} {56}},\ \bibinfo {pages} {3381} (\bibinfo {year} {1997})},\ \Eprint {http://arxiv.org/abs/gr-qc/9610053} {arXiv:gr-qc/9610053} \BibitemShut {NoStop}%
\bibitem [{\citenamefont {Poisson}(1999)}]{Poisson:1999tv}%
  \BibitemOpen
  \bibfield  {author} {\bibinfo {author} {\bibfnamefont {E.}~\bibnamefont {Poisson}},\ }\href@noop {} {\  (\bibinfo {year} {1999})},\ \Eprint {http://arxiv.org/abs/gr-qc/9912045} {arXiv:gr-qc/9912045} \BibitemShut {NoStop}%
\bibitem [{\citenamefont {Gralla}\ \emph {et~al.}(2009)\citenamefont {Gralla}, \citenamefont {Harte},\ and\ \citenamefont {Wald}}]{Gralla:2009md}%
  \BibitemOpen
  \bibfield  {author} {\bibinfo {author} {\bibfnamefont {S.~E.}\ \bibnamefont {Gralla}}, \bibinfo {author} {\bibfnamefont {A.~I.}\ \bibnamefont {Harte}}, \ and\ \bibinfo {author} {\bibfnamefont {R.~M.}\ \bibnamefont {Wald}},\ }\href {\doibase 10.1103/PhysRevD.80.024031} {\bibfield  {journal} {\bibinfo  {journal} {Phys. Rev. D}\ }\textbf {\bibinfo {volume} {80}},\ \bibinfo {pages} {024031} (\bibinfo {year} {2009})},\ \Eprint {http://arxiv.org/abs/0905.2391} {arXiv:0905.2391 [gr-qc]} \BibitemShut {NoStop}%
\bibitem [{\citenamefont {Poisson}\ \emph {et~al.}(2011)\citenamefont {Poisson}, \citenamefont {Pound},\ and\ \citenamefont {Vega}}]{Poisson:2011nh}%
  \BibitemOpen
  \bibfield  {author} {\bibinfo {author} {\bibfnamefont {E.}~\bibnamefont {Poisson}}, \bibinfo {author} {\bibfnamefont {A.}~\bibnamefont {Pound}}, \ and\ \bibinfo {author} {\bibfnamefont {I.}~\bibnamefont {Vega}},\ }\href {\doibase 10.12942/lrr-2011-7} {\bibfield  {journal} {\bibinfo  {journal} {Living Rev. Rel.}\ }\textbf {\bibinfo {volume} {14}},\ \bibinfo {pages} {7} (\bibinfo {year} {2011})},\ \Eprint {http://arxiv.org/abs/1102.0529} {arXiv:1102.0529 [gr-qc]} \BibitemShut {NoStop}%
\bibitem [{\citenamefont {Detweiler}(2012)}]{Detweiler:2011tt}%
  \BibitemOpen
  \bibfield  {author} {\bibinfo {author} {\bibfnamefont {S.}~\bibnamefont {Detweiler}},\ }\href {\doibase 10.1103/PhysRevD.85.044048} {\bibfield  {journal} {\bibinfo  {journal} {Phys. Rev. D}\ }\textbf {\bibinfo {volume} {85}},\ \bibinfo {pages} {044048} (\bibinfo {year} {2012})},\ \Eprint {http://arxiv.org/abs/1107.2098} {arXiv:1107.2098 [gr-qc]} \BibitemShut {NoStop}%
\bibitem [{\citenamefont {Barack}\ and\ \citenamefont {Pound}(2019)}]{Barack:2018yvs}%
  \BibitemOpen
  \bibfield  {author} {\bibinfo {author} {\bibfnamefont {L.}~\bibnamefont {Barack}}\ and\ \bibinfo {author} {\bibfnamefont {A.}~\bibnamefont {Pound}},\ }\href {\doibase 10.1088/1361-6633/aae552} {\bibfield  {journal} {\bibinfo  {journal} {Rept. Prog. Phys.}\ }\textbf {\bibinfo {volume} {82}},\ \bibinfo {pages} {016904} (\bibinfo {year} {2019})},\ \Eprint {http://arxiv.org/abs/1805.10385} {arXiv:1805.10385 [gr-qc]} \BibitemShut {NoStop}%
\bibitem [{\citenamefont {Pound}(2010)}]{Pound:2009sm}%
  \BibitemOpen
  \bibfield  {author} {\bibinfo {author} {\bibfnamefont {A.}~\bibnamefont {Pound}},\ }\href {\doibase 10.1103/PhysRevD.81.024023} {\bibfield  {journal} {\bibinfo  {journal} {Phys. Rev. D}\ }\textbf {\bibinfo {volume} {81}},\ \bibinfo {pages} {024023} (\bibinfo {year} {2010})},\ \Eprint {http://arxiv.org/abs/0907.5197} {arXiv:0907.5197 [gr-qc]} \BibitemShut {NoStop}%
\bibitem [{\citenamefont {Afshordi}\ \emph {et~al.}(2023)\citenamefont {Afshordi} \emph {et~al.}}]{LISAConsortiumWaveformWorkingGroup:2023arg}%
  \BibitemOpen
  \bibfield  {author} {\bibinfo {author} {\bibfnamefont {N.}~\bibnamefont {Afshordi}} \emph {et~al.} (\bibinfo {collaboration} {LISA Consortium Waveform Working Group}),\ }\href@noop {} {\  (\bibinfo {year} {2023})},\ \Eprint {http://arxiv.org/abs/2311.01300} {arXiv:2311.01300 [gr-qc]} \BibitemShut {NoStop}%
\bibitem [{\citenamefont {Breuer}\ \emph {et~al.}(1973)\citenamefont {Breuer}, \citenamefont {Ruffini}, \citenamefont {Tiomno},\ and\ \citenamefont {Vishveshwara}}]{Breuer:1973kt}%
  \BibitemOpen
  \bibfield  {author} {\bibinfo {author} {\bibfnamefont {R.~A.}\ \bibnamefont {Breuer}}, \bibinfo {author} {\bibfnamefont {R.}~\bibnamefont {Ruffini}}, \bibinfo {author} {\bibfnamefont {J.}~\bibnamefont {Tiomno}}, \ and\ \bibinfo {author} {\bibfnamefont {C.~V.}\ \bibnamefont {Vishveshwara}},\ }\href {\doibase 10.1103/PhysRevD.7.1002} {\bibfield  {journal} {\bibinfo  {journal} {Phys. Rev. D}\ }\textbf {\bibinfo {volume} {7}},\ \bibinfo {pages} {1002} (\bibinfo {year} {1973})}\BibitemShut {NoStop}%
\bibitem [{\citenamefont {Detweiler}(1978)}]{Detweiler:1978ge}%
  \BibitemOpen
  \bibfield  {author} {\bibinfo {author} {\bibfnamefont {S.~L.}\ \bibnamefont {Detweiler}},\ }\href {\doibase 10.1086/156529} {\bibfield  {journal} {\bibinfo  {journal} {Astrophys. J.}\ }\textbf {\bibinfo {volume} {225}},\ \bibinfo {pages} {687} (\bibinfo {year} {1978})}\BibitemShut {NoStop}%
\bibitem [{\citenamefont {Gal'tsov}(1982)}]{Galtsov:1982hwm}%
  \BibitemOpen
  \bibfield  {author} {\bibinfo {author} {\bibfnamefont {D.~V.}\ \bibnamefont {Gal'tsov}},\ }\href {\doibase 10.1088/0305-4470/15/12/025} {\bibfield  {journal} {\bibinfo  {journal} {J. Phys. A}\ }\textbf {\bibinfo {volume} {15}},\ \bibinfo {pages} {3737} (\bibinfo {year} {1982})}\BibitemShut {NoStop}%
\bibitem [{\citenamefont {Poisson}(1993)}]{Poisson:1993vp}%
  \BibitemOpen
  \bibfield  {author} {\bibinfo {author} {\bibfnamefont {E.}~\bibnamefont {Poisson}},\ }\href {\doibase 10.1103/PhysRevD.47.1497} {\bibfield  {journal} {\bibinfo  {journal} {Phys. Rev. D}\ }\textbf {\bibinfo {volume} {47}},\ \bibinfo {pages} {1497} (\bibinfo {year} {1993})}\BibitemShut {NoStop}%
\bibitem [{\citenamefont {Detweiler}\ and\ \citenamefont {Whiting}(2003)}]{Detweiler:2002mi}%
  \BibitemOpen
  \bibfield  {author} {\bibinfo {author} {\bibfnamefont {S.~L.}\ \bibnamefont {Detweiler}}\ and\ \bibinfo {author} {\bibfnamefont {B.~F.}\ \bibnamefont {Whiting}},\ }\href {\doibase 10.1103/PhysRevD.67.024025} {\bibfield  {journal} {\bibinfo  {journal} {Phys. Rev. D}\ }\textbf {\bibinfo {volume} {67}},\ \bibinfo {pages} {024025} (\bibinfo {year} {2003})},\ \Eprint {http://arxiv.org/abs/gr-qc/0202086} {arXiv:gr-qc/0202086} \BibitemShut {NoStop}%
\bibitem [{\citenamefont {Pound}(2012)}]{Pound:2012nt}%
  \BibitemOpen
  \bibfield  {author} {\bibinfo {author} {\bibfnamefont {A.}~\bibnamefont {Pound}},\ }\href {\doibase 10.1103/PhysRevLett.109.051101} {\bibfield  {journal} {\bibinfo  {journal} {Phys. Rev. Lett.}\ }\textbf {\bibinfo {volume} {109}},\ \bibinfo {pages} {051101} (\bibinfo {year} {2012})},\ \Eprint {http://arxiv.org/abs/1201.5089} {arXiv:1201.5089 [gr-qc]} \BibitemShut {NoStop}%
\bibitem [{\citenamefont {Torres}\ and\ \citenamefont {Dolan}(2022)}]{Torres:2020fye}%
  \BibitemOpen
  \bibfield  {author} {\bibinfo {author} {\bibfnamefont {T.}~\bibnamefont {Torres}}\ and\ \bibinfo {author} {\bibfnamefont {S.~R.}\ \bibnamefont {Dolan}},\ }\href {\doibase 10.1103/PhysRevD.106.024024} {\bibfield  {journal} {\bibinfo  {journal} {Phys. Rev. D}\ }\textbf {\bibinfo {volume} {106}},\ \bibinfo {pages} {024024} (\bibinfo {year} {2022})},\ \Eprint {http://arxiv.org/abs/2008.12703} {arXiv:2008.12703 [gr-qc]} \BibitemShut {NoStop}%
\bibitem [{\citenamefont {Panosso~Macedo}\ \emph {et~al.}(2022)\citenamefont {Panosso~Macedo}, \citenamefont {Leather}, \citenamefont {Warburton}, \citenamefont {Wardell},\ and\ \citenamefont {Zengino\u{g}lu}}]{PanossoMacedo:2022fdi}%
  \BibitemOpen
  \bibfield  {author} {\bibinfo {author} {\bibfnamefont {R.}~\bibnamefont {Panosso~Macedo}}, \bibinfo {author} {\bibfnamefont {B.}~\bibnamefont {Leather}}, \bibinfo {author} {\bibfnamefont {N.}~\bibnamefont {Warburton}}, \bibinfo {author} {\bibfnamefont {B.}~\bibnamefont {Wardell}}, \ and\ \bibinfo {author} {\bibfnamefont {A.}~\bibnamefont {Zengino\u{g}lu}},\ }\href {\doibase 10.1103/PhysRevD.105.104033} {\bibfield  {journal} {\bibinfo  {journal} {Phys. Rev. D}\ }\textbf {\bibinfo {volume} {105}},\ \bibinfo {pages} {104033} (\bibinfo {year} {2022})},\ \Eprint {http://arxiv.org/abs/2202.01794} {arXiv:2202.01794 [gr-qc]} \BibitemShut {NoStop}%
\bibitem [{\citenamefont {German}\ \emph {et~al.}(2024)\citenamefont {German}, \citenamefont {Cunningham}, \citenamefont {Balakumar}, \citenamefont {Warburton},\ and\ \citenamefont {Dolan}}]{German:2023bye}%
  \BibitemOpen
  \bibfield  {author} {\bibinfo {author} {\bibfnamefont {E.~J.}\ \bibnamefont {German}}, \bibinfo {author} {\bibfnamefont {K.}~\bibnamefont {Cunningham}}, \bibinfo {author} {\bibfnamefont {V.}~\bibnamefont {Balakumar}}, \bibinfo {author} {\bibfnamefont {N.}~\bibnamefont {Warburton}}, \ and\ \bibinfo {author} {\bibfnamefont {S.~R.}\ \bibnamefont {Dolan}},\ }\href {\doibase 10.1103/PhysRevD.109.044068} {\bibfield  {journal} {\bibinfo  {journal} {Phys. Rev. D}\ }\textbf {\bibinfo {volume} {109}},\ \bibinfo {pages} {044068} (\bibinfo {year} {2024})},\ \Eprint {http://arxiv.org/abs/2309.10028} {arXiv:2309.10028 [gr-qc]} \BibitemShut {NoStop}%
\bibitem [{\citenamefont {Miller}\ \emph {et~al.}(2023)\citenamefont {Miller}, \citenamefont {Leather}, \citenamefont {Pound},\ and\ \citenamefont {Warburton}}]{Miller:2023ers}%
  \BibitemOpen
  \bibfield  {author} {\bibinfo {author} {\bibfnamefont {J.}~\bibnamefont {Miller}}, \bibinfo {author} {\bibfnamefont {B.}~\bibnamefont {Leather}}, \bibinfo {author} {\bibfnamefont {A.}~\bibnamefont {Pound}}, \ and\ \bibinfo {author} {\bibfnamefont {N.}~\bibnamefont {Warburton}},\ }\href@noop {} {\  (\bibinfo {year} {2023})},\ \Eprint {http://arxiv.org/abs/2401.00455} {arXiv:2401.00455 [gr-qc]} \BibitemShut {NoStop}%
\bibitem [{\citenamefont {Bourg}\ \emph {et~al.}(2024)\citenamefont {Bourg}, \citenamefont {Leather}, \citenamefont {Casals}, \citenamefont {Pound},\ and\ \citenamefont {Wardell}}]{Bourg:2024vre}%
  \BibitemOpen
  \bibfield  {author} {\bibinfo {author} {\bibfnamefont {P.}~\bibnamefont {Bourg}}, \bibinfo {author} {\bibfnamefont {B.}~\bibnamefont {Leather}}, \bibinfo {author} {\bibfnamefont {M.}~\bibnamefont {Casals}}, \bibinfo {author} {\bibfnamefont {A.}~\bibnamefont {Pound}}, \ and\ \bibinfo {author} {\bibfnamefont {B.}~\bibnamefont {Wardell}},\ }\href@noop {} {\  (\bibinfo {year} {2024})},\ \Eprint {http://arxiv.org/abs/2403.12634} {arXiv:2403.12634 [gr-qc]} \BibitemShut {NoStop}%
\bibitem [{\citenamefont {Bardeen}\ \emph {et~al.}(1972)\citenamefont {Bardeen}, \citenamefont {Press},\ and\ \citenamefont {Teukolsky}}]{Bardeen:1972fi}%
  \BibitemOpen
  \bibfield  {author} {\bibinfo {author} {\bibfnamefont {J.~M.}\ \bibnamefont {Bardeen}}, \bibinfo {author} {\bibfnamefont {W.~H.}\ \bibnamefont {Press}}, \ and\ \bibinfo {author} {\bibfnamefont {S.~A.}\ \bibnamefont {Teukolsky}},\ }\href {\doibase 10.1086/151796} {\bibfield  {journal} {\bibinfo  {journal} {Astrophys. J.}\ }\textbf {\bibinfo {volume} {178}},\ \bibinfo {pages} {347} (\bibinfo {year} {1972})}\BibitemShut {NoStop}%
\bibitem [{\citenamefont {{Piotrovich}}\ \emph {et~al.}(2011)\citenamefont {{Piotrovich}}, \citenamefont {{Silant'ev}}, \citenamefont {{Gnedin}},\ and\ \citenamefont {{Natsvlishvili}}}]{2011AstBu..66..320P}%
  \BibitemOpen
  \bibfield  {author} {\bibinfo {author} {\bibfnamefont {M.~Y.}\ \bibnamefont {{Piotrovich}}}, \bibinfo {author} {\bibfnamefont {N.~A.}\ \bibnamefont {{Silant'ev}}}, \bibinfo {author} {\bibfnamefont {Y.~N.}\ \bibnamefont {{Gnedin}}}, \ and\ \bibinfo {author} {\bibfnamefont {T.~M.}\ \bibnamefont {{Natsvlishvili}}},\ }\href {\doibase 10.1134/S1990341311030047} {\bibfield  {journal} {\bibinfo  {journal} {Astrophysical Bulletin}\ }\textbf {\bibinfo {volume} {66}},\ \bibinfo {pages} {320} (\bibinfo {year} {2011})}\BibitemShut {NoStop}%
\bibitem [{\citenamefont {Eatough}\ \emph {et~al.}(2013)\citenamefont {Eatough} \emph {et~al.}}]{Eatough:2013nva}%
  \BibitemOpen
  \bibfield  {author} {\bibinfo {author} {\bibfnamefont {R.~P.}\ \bibnamefont {Eatough}} \emph {et~al.},\ }\href {\doibase 10.1038/nature12499} {\bibfield  {journal} {\bibinfo  {journal} {Nature}\ }\textbf {\bibinfo {volume} {501}},\ \bibinfo {pages} {391} (\bibinfo {year} {2013})},\ \Eprint {http://arxiv.org/abs/1308.3147} {arXiv:1308.3147 [astro-ph.GA]} \BibitemShut {NoStop}%
\bibitem [{\citenamefont {Baczko}\ \emph {et~al.}(2016)\citenamefont {Baczko} \emph {et~al.}}]{Baczko:2016opl}%
  \BibitemOpen
  \bibfield  {author} {\bibinfo {author} {\bibfnamefont {A.~K.}\ \bibnamefont {Baczko}} \emph {et~al.},\ }\href {\doibase 10.1051/0004-6361/201527951} {\bibfield  {journal} {\bibinfo  {journal} {Astron. Astrophys.}\ }\textbf {\bibinfo {volume} {593}},\ \bibinfo {pages} {A47} (\bibinfo {year} {2016})},\ \Eprint {http://arxiv.org/abs/1605.07100} {arXiv:1605.07100 [astro-ph.GA]} \BibitemShut {NoStop}%
\bibitem [{\citenamefont {Daly}(2019)}]{Daly:2019srb}%
  \BibitemOpen
  \bibfield  {author} {\bibinfo {author} {\bibfnamefont {R.~A.}\ \bibnamefont {Daly}},\ }\href {\doibase 10.3847/1538-4357/ab35e6} {\bibfield  {journal} {\bibinfo  {journal} {Astrophys. J.}\ }\textbf {\bibinfo {volume} {886}},\ \bibinfo {pages} {37} (\bibinfo {year} {2019})},\ \Eprint {http://arxiv.org/abs/1905.11319} {arXiv:1905.11319 [astro-ph.HE]} \BibitemShut {NoStop}%
\bibitem [{\citenamefont {Akiyama}\ \emph {et~al.}(2021)\citenamefont {Akiyama} \emph {et~al.}}]{EventHorizonTelescope:2021srq}%
  \BibitemOpen
  \bibfield  {author} {\bibinfo {author} {\bibfnamefont {K.}~\bibnamefont {Akiyama}} \emph {et~al.} (\bibinfo {collaboration} {Event Horizon Telescope}),\ }\href {\doibase 10.3847/2041-8213/abe4de} {\bibfield  {journal} {\bibinfo  {journal} {Astrophys. J. Lett.}\ }\textbf {\bibinfo {volume} {910}},\ \bibinfo {pages} {L13} (\bibinfo {year} {2021})},\ \Eprint {http://arxiv.org/abs/2105.01173} {arXiv:2105.01173 [astro-ph.HE]} \BibitemShut {NoStop}%
\bibitem [{\citenamefont {Frolov}\ and\ \citenamefont {Shoom}(2010)}]{Frolov:2010mi}%
  \BibitemOpen
  \bibfield  {author} {\bibinfo {author} {\bibfnamefont {V.~P.}\ \bibnamefont {Frolov}}\ and\ \bibinfo {author} {\bibfnamefont {A.~A.}\ \bibnamefont {Shoom}},\ }\href {\doibase 10.1103/PhysRevD.82.084034} {\bibfield  {journal} {\bibinfo  {journal} {Phys. Rev. D}\ }\textbf {\bibinfo {volume} {82}},\ \bibinfo {pages} {084034} (\bibinfo {year} {2010})},\ \Eprint {http://arxiv.org/abs/1008.2985} {arXiv:1008.2985 [gr-qc]} \BibitemShut {NoStop}%
\bibitem [{\citenamefont {Frolov}(2012)}]{Frolov:2011ea}%
  \BibitemOpen
  \bibfield  {author} {\bibinfo {author} {\bibfnamefont {V.~P.}\ \bibnamefont {Frolov}},\ }\href {\doibase 10.1103/PhysRevD.85.024020} {\bibfield  {journal} {\bibinfo  {journal} {Phys. Rev. D}\ }\textbf {\bibinfo {volume} {85}},\ \bibinfo {pages} {024020} (\bibinfo {year} {2012})},\ \Eprint {http://arxiv.org/abs/1110.6274} {arXiv:1110.6274 [gr-qc]} \BibitemShut {NoStop}%
\bibitem [{\citenamefont {Zahrani}\ \emph {et~al.}(2013)\citenamefont {Zahrani}, \citenamefont {Frolov},\ and\ \citenamefont {Shoom}}]{Zahrani:2013up}%
  \BibitemOpen
  \bibfield  {author} {\bibinfo {author} {\bibfnamefont {A.~M.~A.}\ \bibnamefont {Zahrani}}, \bibinfo {author} {\bibfnamefont {V.~P.}\ \bibnamefont {Frolov}}, \ and\ \bibinfo {author} {\bibfnamefont {A.~A.}\ \bibnamefont {Shoom}},\ }\href {\doibase 10.1103/PhysRevD.87.084043} {\bibfield  {journal} {\bibinfo  {journal} {Phys. Rev. D}\ }\textbf {\bibinfo {volume} {87}},\ \bibinfo {pages} {084043} (\bibinfo {year} {2013})},\ \Eprint {http://arxiv.org/abs/1301.4633} {arXiv:1301.4633 [gr-qc]} \BibitemShut {NoStop}%
\bibitem [{\citenamefont {Kolo\v{s}}\ \emph {et~al.}(2015)\citenamefont {Kolo\v{s}}, \citenamefont {Stuchl\'\i{}k},\ and\ \citenamefont {Tursunov}}]{Kolos:2015iva}%
  \BibitemOpen
  \bibfield  {author} {\bibinfo {author} {\bibfnamefont {M.}~\bibnamefont {Kolo\v{s}}}, \bibinfo {author} {\bibfnamefont {Z.}~\bibnamefont {Stuchl\'\i{}k}}, \ and\ \bibinfo {author} {\bibfnamefont {A.}~\bibnamefont {Tursunov}},\ }\href {\doibase 10.1088/0264-9381/32/16/165009} {\bibfield  {journal} {\bibinfo  {journal} {Class. Quant. Grav.}\ }\textbf {\bibinfo {volume} {32}},\ \bibinfo {pages} {165009} (\bibinfo {year} {2015})},\ \Eprint {http://arxiv.org/abs/1506.06799} {arXiv:1506.06799 [gr-qc]} \BibitemShut {NoStop}%
\bibitem [{\citenamefont {Qi}\ \emph {et~al.}(2023)\citenamefont {Qi}, \citenamefont {Rayimbaev},\ and\ \citenamefont {Ahmedov}}]{Qi:2023brf}%
  \BibitemOpen
  \bibfield  {author} {\bibinfo {author} {\bibfnamefont {M.}~\bibnamefont {Qi}}, \bibinfo {author} {\bibfnamefont {J.}~\bibnamefont {Rayimbaev}}, \ and\ \bibinfo {author} {\bibfnamefont {B.}~\bibnamefont {Ahmedov}},\ }\href {\doibase 10.1140/epjc/s10052-023-11912-1} {\bibfield  {journal} {\bibinfo  {journal} {Eur. Phys. J. C}\ }\textbf {\bibinfo {volume} {83}},\ \bibinfo {pages} {730} (\bibinfo {year} {2023})}\BibitemShut {NoStop}%
\bibitem [{\citenamefont {Baker}\ and\ \citenamefont {Frolov}(2023)}]{Baker:2023gdc}%
  \BibitemOpen
  \bibfield  {author} {\bibinfo {author} {\bibfnamefont {N.~P.}\ \bibnamefont {Baker}}\ and\ \bibinfo {author} {\bibfnamefont {V.~P.}\ \bibnamefont {Frolov}},\ }\href@noop {} {\  (\bibinfo {year} {2023})},\ \Eprint {http://arxiv.org/abs/2305.12591} {arXiv:2305.12591 [gr-qc]} \BibitemShut {NoStop}%
\bibitem [{\citenamefont {Aliev}\ and\ \citenamefont {Galtsov}(1981)}]{Aliev:1980hz}%
  \BibitemOpen
  \bibfield  {author} {\bibinfo {author} {\bibfnamefont {A.~N.}\ \bibnamefont {Aliev}}\ and\ \bibinfo {author} {\bibfnamefont {D.~V.}\ \bibnamefont {Galtsov}},\ }\href {\doibase 10.1007/BF00756068} {\bibfield  {journal} {\bibinfo  {journal} {Gen. Rel. Grav.}\ }\textbf {\bibinfo {volume} {13}},\ \bibinfo {pages} {899} (\bibinfo {year} {1981})}\BibitemShut {NoStop}%
\bibitem [{\citenamefont {Sokolov}\ \emph {et~al.}(1978)\citenamefont {Sokolov}, \citenamefont {Galtsov},\ and\ \citenamefont {Petukhov}}]{Sokolov:1978tc}%
  \BibitemOpen
  \bibfield  {author} {\bibinfo {author} {\bibfnamefont {A.~A.}\ \bibnamefont {Sokolov}}, \bibinfo {author} {\bibfnamefont {D.~V.}\ \bibnamefont {Galtsov}}, \ and\ \bibinfo {author} {\bibfnamefont {V.~I.}\ \bibnamefont {Petukhov}},\ }\href {\doibase 10.1016/0375-9601(78)90737-5} {\bibfield  {journal} {\bibinfo  {journal} {Phys. Lett. A}\ }\textbf {\bibinfo {volume} {68}},\ \bibinfo {pages} {1} (\bibinfo {year} {1978})}\BibitemShut {NoStop}%
\bibitem [{\citenamefont {{Tursunov}}\ \emph {et~al.}(2018)\citenamefont {{Tursunov}}, \citenamefont {{Kolo{\v{s}}}}, \citenamefont {{Stuchl{\'\i}k}},\ and\ \citenamefont {{Gal'tsov}}}]{2018ApJ...861....2T}%
  \BibitemOpen
  \bibfield  {author} {\bibinfo {author} {\bibfnamefont {A.}~\bibnamefont {{Tursunov}}}, \bibinfo {author} {\bibfnamefont {M.}~\bibnamefont {{Kolo{\v{s}}}}}, \bibinfo {author} {\bibfnamefont {Z.}~\bibnamefont {{Stuchl{\'\i}k}}}, \ and\ \bibinfo {author} {\bibfnamefont {D.~V.}\ \bibnamefont {{Gal'tsov}}},\ }\href {\doibase 10.3847/1538-4357/aac7c5} {\bibfield  {journal} {\bibinfo  {journal} {The Astrophysical Journal}\ }\textbf {\bibinfo {volume} {861}},\ \bibinfo {eid} {2} (\bibinfo {year} {2018})},\ \Eprint {http://arxiv.org/abs/1803.09682} {arXiv:1803.09682 [gr-qc]} \BibitemShut {NoStop}%
\bibitem [{\citenamefont {Wald}(1974)}]{PhysRevD.10.1680}%
  \BibitemOpen
  \bibfield  {author} {\bibinfo {author} {\bibfnamefont {R.~M.}\ \bibnamefont {Wald}},\ }\href {\doibase 10.1103/PhysRevD.10.1680} {\bibfield  {journal} {\bibinfo  {journal} {Phys. Rev. D}\ }\textbf {\bibinfo {volume} {10}},\ \bibinfo {pages} {1680} (\bibinfo {year} {1974})}\BibitemShut {NoStop}%
\bibitem [{\citenamefont {Galtsov}\ and\ \citenamefont {Petukhov}(1978)}]{Galtsov:1978ag}%
  \BibitemOpen
  \bibfield  {author} {\bibinfo {author} {\bibfnamefont {D.~V.}\ \bibnamefont {Galtsov}}\ and\ \bibinfo {author} {\bibfnamefont {V.~I.}\ \bibnamefont {Petukhov}},\ }\href@noop {} {\bibfield  {journal} {\bibinfo  {journal} {Zh. Eksp. Teor. Fiz.}\ }\textbf {\bibinfo {volume} {74}},\ \bibinfo {pages} {801} (\bibinfo {year} {1978})}\BibitemShut {NoStop}%
\bibitem [{\citenamefont {Tursunov}\ \emph {et~al.}(2018)\citenamefont {Tursunov}, \citenamefont {Kolo\v{s}},\ and\ \citenamefont {Stuchl\'\i{}k}}]{Tursunov:2018udx}%
  \BibitemOpen
  \bibfield  {author} {\bibinfo {author} {\bibfnamefont {A.}~\bibnamefont {Tursunov}}, \bibinfo {author} {\bibfnamefont {M.}~\bibnamefont {Kolo\v{s}}}, \ and\ \bibinfo {author} {\bibfnamefont {Z.}~\bibnamefont {Stuchl\'\i{}k}},\ }\href {\doibase 10.1002/asna.201813502} {\bibfield  {journal} {\bibinfo  {journal} {Astron. Nachr.}\ }\textbf {\bibinfo {volume} {339}},\ \bibinfo {pages} {341} (\bibinfo {year} {2018})},\ \Eprint {http://arxiv.org/abs/1806.06754} {arXiv:1806.06754 [gr-qc]} \BibitemShut {NoStop}%
\bibitem [{\citenamefont {Juraev}\ \emph {et~al.}(2024)\citenamefont {Juraev}, \citenamefont {Stuchl\'\i{}k}, \citenamefont {Tursunov},\ and\ \citenamefont {Kolo\v{s}}}]{Juraev:2024dju}%
  \BibitemOpen
  \bibfield  {author} {\bibinfo {author} {\bibfnamefont {B.}~\bibnamefont {Juraev}}, \bibinfo {author} {\bibfnamefont {Z.}~\bibnamefont {Stuchl\'\i{}k}}, \bibinfo {author} {\bibfnamefont {A.}~\bibnamefont {Tursunov}}, \ and\ \bibinfo {author} {\bibfnamefont {M.}~\bibnamefont {Kolo\v{s}}},\ }\href@noop {} {\  (\bibinfo {year} {2024})},\ \Eprint {http://arxiv.org/abs/2402.13797} {arXiv:2402.13797 [gr-qc]} \BibitemShut {NoStop}%
\bibitem [{\citenamefont {Santos}\ \emph {et~al.}(2023)\citenamefont {Santos}, \citenamefont {Cardoso},\ and\ \citenamefont {Nat\'ario}}]{Santos:2023uka}%
  \BibitemOpen
  \bibfield  {author} {\bibinfo {author} {\bibfnamefont {J.~S.}\ \bibnamefont {Santos}}, \bibinfo {author} {\bibfnamefont {V.}~\bibnamefont {Cardoso}}, \ and\ \bibinfo {author} {\bibfnamefont {J.}~\bibnamefont {Nat\'ario}},\ }\href {\doibase 10.1103/PhysRevD.107.064046} {\bibfield  {journal} {\bibinfo  {journal} {Phys. Rev. D}\ }\textbf {\bibinfo {volume} {107}},\ \bibinfo {pages} {064046} (\bibinfo {year} {2023})},\ \Eprint {http://arxiv.org/abs/2303.03411} {arXiv:2303.03411 [gr-qc]} \BibitemShut {NoStop}%
\bibitem [{\citenamefont {Gron}\ and\ \citenamefont {Nass}(2008)}]{Gron:2008tr}%
  \BibitemOpen
  \bibfield  {author} {\bibinfo {author} {\bibfnamefont {O.}~\bibnamefont {Gron}}\ and\ \bibinfo {author} {\bibfnamefont {S.~K.}\ \bibnamefont {Nass}},\ }\href@noop {} {\  (\bibinfo {year} {2008})},\ \Eprint {http://arxiv.org/abs/0806.0464} {arXiv:0806.0464 [gr-qc]} \BibitemShut {NoStop}%
\bibitem [{\citenamefont {Brito}\ \emph {et~al.}(2012)\citenamefont {Brito}, \citenamefont {Cardoso},\ and\ \citenamefont {Pani}}]{Brito:2012gw}%
  \BibitemOpen
  \bibfield  {author} {\bibinfo {author} {\bibfnamefont {R.}~\bibnamefont {Brito}}, \bibinfo {author} {\bibfnamefont {V.}~\bibnamefont {Cardoso}}, \ and\ \bibinfo {author} {\bibfnamefont {P.}~\bibnamefont {Pani}},\ }\href {\doibase 10.1103/PhysRevD.86.024032} {\bibfield  {journal} {\bibinfo  {journal} {Phys. Rev. D}\ }\textbf {\bibinfo {volume} {86}},\ \bibinfo {pages} {024032} (\bibinfo {year} {2012})},\ \Eprint {http://arxiv.org/abs/1207.0504} {arXiv:1207.0504 [gr-qc]} \BibitemShut {NoStop}%
\bibitem [{\citenamefont {Jackson}(1998)}]{Jackson:1998nia}%
  \BibitemOpen
  \bibfield  {author} {\bibinfo {author} {\bibfnamefont {J.~D.}\ \bibnamefont {Jackson}},\ }\href@noop {} {\emph {\bibinfo {title} {{Classical Electrodynamics}}}}\ (\bibinfo  {publisher} {Wiley},\ \bibinfo {year} {1998})\BibitemShut {NoStop}%
\bibitem [{\citenamefont {Flanagan}\ and\ \citenamefont {Wald}(1996)}]{Flanagan:1996gw}%
  \BibitemOpen
  \bibfield  {author} {\bibinfo {author} {\bibfnamefont {E.~E.}\ \bibnamefont {Flanagan}}\ and\ \bibinfo {author} {\bibfnamefont {R.~M.}\ \bibnamefont {Wald}},\ }\href {\doibase 10.1103/PhysRevD.54.6233} {\bibfield  {journal} {\bibinfo  {journal} {Phys. Rev. D}\ }\textbf {\bibinfo {volume} {54}},\ \bibinfo {pages} {6233} (\bibinfo {year} {1996})},\ \Eprint {http://arxiv.org/abs/gr-qc/9602052} {arXiv:gr-qc/9602052} \BibitemShut {NoStop}%
\bibitem [{\citenamefont {Spohn}(2000)}]{Spohn:1999uf}%
  \BibitemOpen
  \bibfield  {author} {\bibinfo {author} {\bibfnamefont {H.}~\bibnamefont {Spohn}},\ }\href {\doibase 10.1209/epl/i2000-00268-x} {\bibfield  {journal} {\bibinfo  {journal} {EPL}\ }\textbf {\bibinfo {volume} {50}},\ \bibinfo {pages} {287} (\bibinfo {year} {2000})},\ \Eprint {http://arxiv.org/abs/physics/9911027} {arXiv:physics/9911027} \BibitemShut {NoStop}%
\bibitem [{\citenamefont {{Rohrlich}}(2001)}]{2001PhLA..283..276R}%
  \BibitemOpen
  \bibfield  {author} {\bibinfo {author} {\bibfnamefont {F.}~\bibnamefont {{Rohrlich}}},\ }\href {\doibase 10.1016/S0375-9601(01)00264-X} {\bibfield  {journal} {\bibinfo  {journal} {Physics Letters A}\ }\textbf {\bibinfo {volume} {283}},\ \bibinfo {pages} {276} (\bibinfo {year} {2001})}\BibitemShut {NoStop}%
\bibitem [{\citenamefont {DeWitt}\ and\ \citenamefont {DeWitt}(1964)}]{DeWitt:1964de}%
  \BibitemOpen
  \bibfield  {author} {\bibinfo {author} {\bibfnamefont {C.~M.}\ \bibnamefont {DeWitt}}\ and\ \bibinfo {author} {\bibfnamefont {B.~S.}\ \bibnamefont {DeWitt}},\ }\href {\doibase 10.1103/PhysicsPhysiqueFizika.1.3} {\bibfield  {journal} {\bibinfo  {journal} {Physics Physique Fizika}\ }\textbf {\bibinfo {volume} {1}},\ \bibinfo {pages} {3} (\bibinfo {year} {1964})},\ \bibinfo {note} {[Erratum: Physics Physique Fizika 1, 145 (1964)]}\BibitemShut {NoStop}%
\bibitem [{\citenamefont {Teukolsky}(1972)}]{Teukolsky:1972my}%
  \BibitemOpen
  \bibfield  {author} {\bibinfo {author} {\bibfnamefont {S.~A.}\ \bibnamefont {Teukolsky}},\ }\href {\doibase 10.1103/PhysRevLett.29.1114} {\bibfield  {journal} {\bibinfo  {journal} {Phys. Rev. Lett.}\ }\textbf {\bibinfo {volume} {29}},\ \bibinfo {pages} {1114} (\bibinfo {year} {1972})}\BibitemShut {NoStop}%
\bibitem [{\citenamefont {Teukolsky}(1973)}]{Teukolsky:1973ha}%
  \BibitemOpen
  \bibfield  {author} {\bibinfo {author} {\bibfnamefont {S.~A.}\ \bibnamefont {Teukolsky}},\ }\href {\doibase 10.1086/152444} {\bibfield  {journal} {\bibinfo  {journal} {Astrophys. J.}\ }\textbf {\bibinfo {volume} {185}},\ \bibinfo {pages} {635} (\bibinfo {year} {1973})}\BibitemShut {NoStop}%
\bibitem [{\citenamefont {Bardeen}\ and\ \citenamefont {Press}(1973)}]{Bardeen:1973xb}%
  \BibitemOpen
  \bibfield  {author} {\bibinfo {author} {\bibfnamefont {J.~M.}\ \bibnamefont {Bardeen}}\ and\ \bibinfo {author} {\bibfnamefont {W.~H.}\ \bibnamefont {Press}},\ }\href {\doibase 10.1063/1.1666175} {\bibfield  {journal} {\bibinfo  {journal} {J. Math. Phys.}\ }\textbf {\bibinfo {volume} {14}},\ \bibinfo {pages} {7} (\bibinfo {year} {1973})}\BibitemShut {NoStop}%
\bibitem [{\citenamefont {Newman}\ and\ \citenamefont {Penrose}(1962)}]{doi:10.1063/1.1724257}%
  \BibitemOpen
  \bibfield  {author} {\bibinfo {author} {\bibfnamefont {E.}~\bibnamefont {Newman}}\ and\ \bibinfo {author} {\bibfnamefont {R.}~\bibnamefont {Penrose}},\ }\href {\doibase 10.1063/1.1724257} {\bibfield  {journal} {\bibinfo  {journal} {Journal of Mathematical Physics}\ }\textbf {\bibinfo {volume} {3}},\ \bibinfo {pages} {566} (\bibinfo {year} {1962})},\ \Eprint {http://arxiv.org/abs/https://doi.org/10.1063/1.1724257} {https://doi.org/10.1063/1.1724257} \BibitemShut {NoStop}%
\bibitem [{\citenamefont {{Goldberg}}\ \emph {et~al.}(1967)\citenamefont {{Goldberg}}, \citenamefont {{Macfarlane}}, \citenamefont {{Newman}}, \citenamefont {{Rohrlich}},\ and\ \citenamefont {{Sudarshan}}}]{1967JMP.....8.2155G}%
  \BibitemOpen
  \bibfield  {author} {\bibinfo {author} {\bibfnamefont {J.~N.}\ \bibnamefont {{Goldberg}}}, \bibinfo {author} {\bibfnamefont {A.~J.}\ \bibnamefont {{Macfarlane}}}, \bibinfo {author} {\bibfnamefont {E.~T.}\ \bibnamefont {{Newman}}}, \bibinfo {author} {\bibfnamefont {F.}~\bibnamefont {{Rohrlich}}}, \ and\ \bibinfo {author} {\bibfnamefont {E.~C.~G.}\ \bibnamefont {{Sudarshan}}},\ }\href {\doibase 10.1063/1.1705135} {\bibfield  {journal} {\bibinfo  {journal} {Journal of Mathematical Physics}\ }\textbf {\bibinfo {volume} {8}},\ \bibinfo {pages} {2155} (\bibinfo {year} {1967})}\BibitemShut {NoStop}%
\bibitem [{\citenamefont {Teukolsky}\ and\ \citenamefont {Press}(1974)}]{Teukolsky:1974yv}%
  \BibitemOpen
  \bibfield  {author} {\bibinfo {author} {\bibfnamefont {S.~A.}\ \bibnamefont {Teukolsky}}\ and\ \bibinfo {author} {\bibfnamefont {W.~H.}\ \bibnamefont {Press}},\ }\href {\doibase 10.1086/153180} {\bibfield  {journal} {\bibinfo  {journal} {Astrophys. J.}\ }\textbf {\bibinfo {volume} {193}},\ \bibinfo {pages} {443} (\bibinfo {year} {1974})}\BibitemShut {NoStop}%
\bibitem [{\citenamefont {Bekenstein}(1973)}]{Bekenstein:1973mi}%
  \BibitemOpen
  \bibfield  {author} {\bibinfo {author} {\bibfnamefont {J.~D.}\ \bibnamefont {Bekenstein}},\ }\href {\doibase 10.1103/PhysRevD.7.949} {\bibfield  {journal} {\bibinfo  {journal} {Phys. Rev. D}\ }\textbf {\bibinfo {volume} {7}},\ \bibinfo {pages} {949} (\bibinfo {year} {1973})}\BibitemShut {NoStop}%
\bibitem [{\citenamefont {Cardoso}\ \emph {et~al.}(2021{\natexlab{a}})\citenamefont {Cardoso}, \citenamefont {Macedo},\ and\ \citenamefont {Vicente}}]{Cardoso:2020iji}%
  \BibitemOpen
  \bibfield  {author} {\bibinfo {author} {\bibfnamefont {V.}~\bibnamefont {Cardoso}}, \bibinfo {author} {\bibfnamefont {C.~F.~B.}\ \bibnamefont {Macedo}}, \ and\ \bibinfo {author} {\bibfnamefont {R.}~\bibnamefont {Vicente}},\ }\href {\doibase 10.1103/PhysRevD.103.023015} {\bibfield  {journal} {\bibinfo  {journal} {Phys. Rev. D}\ }\textbf {\bibinfo {volume} {103}},\ \bibinfo {pages} {023015} (\bibinfo {year} {2021}{\natexlab{a}})},\ \Eprint {http://arxiv.org/abs/2010.15151} {arXiv:2010.15151 [gr-qc]} \BibitemShut {NoStop}%
\bibitem [{\citenamefont {Cardoso}\ \emph {et~al.}(2019)\citenamefont {Cardoso}, \citenamefont {del Rio},\ and\ \citenamefont {Kimura}}]{Cardoso:2019nis}%
  \BibitemOpen
  \bibfield  {author} {\bibinfo {author} {\bibfnamefont {V.}~\bibnamefont {Cardoso}}, \bibinfo {author} {\bibfnamefont {A.}~\bibnamefont {del Rio}}, \ and\ \bibinfo {author} {\bibfnamefont {M.}~\bibnamefont {Kimura}},\ }\href {\doibase 10.1103/PhysRevD.100.084046} {\bibfield  {journal} {\bibinfo  {journal} {Phys. Rev. D}\ }\textbf {\bibinfo {volume} {100}},\ \bibinfo {pages} {084046} (\bibinfo {year} {2019})},\ \bibinfo {note} {[Erratum: Phys.Rev.D 101, 069902 (2020)]},\ \Eprint {http://arxiv.org/abs/1907.01561} {arXiv:1907.01561 [gr-qc]} \BibitemShut {NoStop}%
\bibitem [{\citenamefont {Poisson}\ and\ \citenamefont {Sasaki}(1995)}]{Poisson:1994yf}%
  \BibitemOpen
  \bibfield  {author} {\bibinfo {author} {\bibfnamefont {E.}~\bibnamefont {Poisson}}\ and\ \bibinfo {author} {\bibfnamefont {M.}~\bibnamefont {Sasaki}},\ }\href {\doibase 10.1103/PhysRevD.51.5753} {\bibfield  {journal} {\bibinfo  {journal} {Phys. Rev. D}\ }\textbf {\bibinfo {volume} {51}},\ \bibinfo {pages} {5753} (\bibinfo {year} {1995})},\ \Eprint {http://arxiv.org/abs/gr-qc/9412027} {arXiv:gr-qc/9412027} \BibitemShut {NoStop}%
\bibitem [{\citenamefont {Cardoso}\ \emph {et~al.}(2021{\natexlab{b}})\citenamefont {Cardoso}, \citenamefont {Duque},\ and\ \citenamefont {Khanna}}]{Cardoso:2021vjq}%
  \BibitemOpen
  \bibfield  {author} {\bibinfo {author} {\bibfnamefont {V.}~\bibnamefont {Cardoso}}, \bibinfo {author} {\bibfnamefont {F.}~\bibnamefont {Duque}}, \ and\ \bibinfo {author} {\bibfnamefont {G.}~\bibnamefont {Khanna}},\ }\href {\doibase 10.1103/PhysRevD.103.L081501} {\bibfield  {journal} {\bibinfo  {journal} {Phys. Rev. D}\ }\textbf {\bibinfo {volume} {103}},\ \bibinfo {pages} {L081501} (\bibinfo {year} {2021}{\natexlab{b}})},\ \Eprint {http://arxiv.org/abs/2101.01186} {arXiv:2101.01186 [gr-qc]} \BibitemShut {NoStop}%
\bibitem [{\citenamefont {Cardoso}\ and\ \citenamefont {Duque}(2022)}]{Cardoso:2022fbq}%
  \BibitemOpen
  \bibfield  {author} {\bibinfo {author} {\bibfnamefont {V.}~\bibnamefont {Cardoso}}\ and\ \bibinfo {author} {\bibfnamefont {F.}~\bibnamefont {Duque}},\ }\href {\doibase 10.1103/PhysRevD.105.104023} {\bibfield  {journal} {\bibinfo  {journal} {Phys. Rev. D}\ }\textbf {\bibinfo {volume} {105}},\ \bibinfo {pages} {104023} (\bibinfo {year} {2022})},\ \Eprint {http://arxiv.org/abs/2204.05315} {arXiv:2204.05315 [gr-qc]} \BibitemShut {NoStop}%
\bibitem [{\citenamefont {Cardoso}\ \emph {et~al.}(2022)\citenamefont {Cardoso}, \citenamefont {Destounis}, \citenamefont {Duque}, \citenamefont {Panosso~Macedo},\ and\ \citenamefont {Maselli}}]{Cardoso:2022whc}%
  \BibitemOpen
  \bibfield  {author} {\bibinfo {author} {\bibfnamefont {V.}~\bibnamefont {Cardoso}}, \bibinfo {author} {\bibfnamefont {K.}~\bibnamefont {Destounis}}, \bibinfo {author} {\bibfnamefont {F.}~\bibnamefont {Duque}}, \bibinfo {author} {\bibfnamefont {R.}~\bibnamefont {Panosso~Macedo}}, \ and\ \bibinfo {author} {\bibfnamefont {A.}~\bibnamefont {Maselli}},\ }\href {\doibase 10.1103/PhysRevLett.129.241103} {\bibfield  {journal} {\bibinfo  {journal} {Phys. Rev. Lett.}\ }\textbf {\bibinfo {volume} {129}},\ \bibinfo {pages} {241103} (\bibinfo {year} {2022})},\ \Eprint {http://arxiv.org/abs/2210.01133} {arXiv:2210.01133 [gr-qc]} \BibitemShut {NoStop}%
\bibitem [{\citenamefont {Cardoso}\ \emph {et~al.}(2011)\citenamefont {Cardoso}, \citenamefont {Chakrabarti}, \citenamefont {Pani}, \citenamefont {Berti},\ and\ \citenamefont {Gualtieri}}]{Cardoso:2011xi}%
  \BibitemOpen
  \bibfield  {author} {\bibinfo {author} {\bibfnamefont {V.}~\bibnamefont {Cardoso}}, \bibinfo {author} {\bibfnamefont {S.}~\bibnamefont {Chakrabarti}}, \bibinfo {author} {\bibfnamefont {P.}~\bibnamefont {Pani}}, \bibinfo {author} {\bibfnamefont {E.}~\bibnamefont {Berti}}, \ and\ \bibinfo {author} {\bibfnamefont {L.}~\bibnamefont {Gualtieri}},\ }\href {\doibase 10.1103/PhysRevLett.107.241101} {\bibfield  {journal} {\bibinfo  {journal} {Phys. Rev. Lett.}\ }\textbf {\bibinfo {volume} {107}},\ \bibinfo {pages} {241101} (\bibinfo {year} {2011})},\ \Eprint {http://arxiv.org/abs/1109.6021} {arXiv:1109.6021 [gr-qc]} \BibitemShut {NoStop}%
\bibitem [{\citenamefont {Starobinskil}\ and\ \citenamefont {Churilov}(1974)}]{Starobinskil:1974nkd}%
  \BibitemOpen
  \bibfield  {author} {\bibinfo {author} {\bibfnamefont {A.~A.}\ \bibnamefont {Starobinskil}}\ and\ \bibinfo {author} {\bibfnamefont {S.~M.}\ \bibnamefont {Churilov}},\ }\href@noop {} {\bibfield  {journal} {\bibinfo  {journal} {Sov. Phys. JETP}\ }\textbf {\bibinfo {volume} {65}},\ \bibinfo {pages} {1} (\bibinfo {year} {1974})}\BibitemShut {NoStop}%
\bibitem [{\citenamefont {Page}(1976)}]{Page:1976df}%
  \BibitemOpen
  \bibfield  {author} {\bibinfo {author} {\bibfnamefont {D.~N.}\ \bibnamefont {Page}},\ }\href {\doibase 10.1103/PhysRevD.13.198} {\bibfield  {journal} {\bibinfo  {journal} {Phys. Rev. D}\ }\textbf {\bibinfo {volume} {13}},\ \bibinfo {pages} {198} (\bibinfo {year} {1976})}\BibitemShut {NoStop}%
\bibitem [{\citenamefont {{Leaver}}(1986)}]{1986JMP....27.1238L}%
  \BibitemOpen
  \bibfield  {author} {\bibinfo {author} {\bibfnamefont {E.~W.}\ \bibnamefont {{Leaver}}},\ }\href {\doibase 10.1063/1.527130} {\bibfield  {journal} {\bibinfo  {journal} {Journal of Mathematical Physics}\ }\textbf {\bibinfo {volume} {27}},\ \bibinfo {pages} {1238} (\bibinfo {year} {1986})}\BibitemShut {NoStop}%
\bibitem [{\citenamefont {Mano}\ \emph {et~al.}(1996)\citenamefont {Mano}, \citenamefont {Suzuki},\ and\ \citenamefont {Takasugi}}]{Mano:1996vt}%
  \BibitemOpen
  \bibfield  {author} {\bibinfo {author} {\bibfnamefont {S.}~\bibnamefont {Mano}}, \bibinfo {author} {\bibfnamefont {H.}~\bibnamefont {Suzuki}}, \ and\ \bibinfo {author} {\bibfnamefont {E.}~\bibnamefont {Takasugi}},\ }\href {\doibase 10.1143/PTP.95.1079} {\bibfield  {journal} {\bibinfo  {journal} {Prog. Theor. Phys.}\ }\textbf {\bibinfo {volume} {95}},\ \bibinfo {pages} {1079} (\bibinfo {year} {1996})},\ \Eprint {http://arxiv.org/abs/gr-qc/9603020} {arXiv:gr-qc/9603020} \BibitemShut {NoStop}%
\bibitem [{\citenamefont {Bayliss}\ and\ \citenamefont {Turkel}(1980)}]{Bayliss:1980}%
  \BibitemOpen
  \bibfield  {author} {\bibinfo {author} {\bibfnamefont {A.}~\bibnamefont {Bayliss}}\ and\ \bibinfo {author} {\bibfnamefont {E.}~\bibnamefont {Turkel}},\ }\href {\doibase https://doi.org/10.1002/cpa.3160330603} {\bibfield  {journal} {\bibinfo  {journal} {Communications on Pure and Applied Mathematics}\ }\textbf {\bibinfo {volume} {33}},\ \bibinfo {pages} {707} (\bibinfo {year} {1980})},\ \Eprint {http://arxiv.org/abs/https://onlinelibrary.wiley.com/doi/pdf/10.1002/cpa.3160330603} {https://onlinelibrary.wiley.com/doi/pdf/10.1002/cpa.3160330603} \BibitemShut {NoStop}%
\bibitem [{\citenamefont {Sarbach}(2007)}]{Sarbach:2007hd}%
  \BibitemOpen
  \bibfield  {author} {\bibinfo {author} {\bibfnamefont {O.}~\bibnamefont {Sarbach}},\ }\href {\doibase 10.1088/1742-6596/91/1/012005} {\bibfield  {journal} {\bibinfo  {journal} {J. Phys. Conf. Ser.}\ }\textbf {\bibinfo {volume} {91}},\ \bibinfo {pages} {012005} (\bibinfo {year} {2007})},\ \Eprint {http://arxiv.org/abs/0708.4266} {arXiv:0708.4266 [gr-qc]} \BibitemShut {NoStop}%
\bibitem [{\citenamefont {Yunes}\ \emph {et~al.}(2012)\citenamefont {Yunes}, \citenamefont {Pani},\ and\ \citenamefont {Cardoso}}]{Yunes:2011aa}%
  \BibitemOpen
  \bibfield  {author} {\bibinfo {author} {\bibfnamefont {N.}~\bibnamefont {Yunes}}, \bibinfo {author} {\bibfnamefont {P.}~\bibnamefont {Pani}}, \ and\ \bibinfo {author} {\bibfnamefont {V.}~\bibnamefont {Cardoso}},\ }\href {\doibase 10.1103/PhysRevD.85.102003} {\bibfield  {journal} {\bibinfo  {journal} {Phys. Rev. D}\ }\textbf {\bibinfo {volume} {85}},\ \bibinfo {pages} {102003} (\bibinfo {year} {2012})},\ \Eprint {http://arxiv.org/abs/1112.3351} {arXiv:1112.3351 [gr-qc]} \BibitemShut {NoStop}%
\end{thebibliography}
\end{document}